\documentclass[arps,prd,onecolumn,showpacs]{revtex4}
\usepackage{graphicx}
\usepackage{amsmath}
\usepackage{amsfonts}
\usepackage{amssymb}
\usepackage{color}
\usepackage{epsf}

\newcommand{\be}{\begin{equation}}         
\newcommand{\ee}{\end{equation}}
\newcommand{\ba}{\begin{eqnarray}}
\newcommand{\ea}{\end{eqnarray}}

\newcommand\lsim{\mathrel{\rlap{\lower4pt\hbox{\hskip1pt$\sim$}}
        \raise1pt\hbox{$<$}}}
\newcommand\gsim{\mathrel{\rlap{\lower4pt\hbox{\hskip1pt$\sim$}}
        \raise1pt\hbox{$>$}}}
\def\k{{\bf k}}
\def\x{{\bf x}}

\begin{document}

\title{Halo bias in mixed dark matter cosmologies}
\author{ Marilena LoVerde}
\affiliation{ Enrico Fermi Institute, Kavli Institute for Cosmological Physics, Department of Astronomy and Astrophysics, University of Chicago, Illinois, 60637, U.S.A.}
\begin{abstract}
The large-scale distribution of cold dark matter halos is generally assumed to trace the large-scale distribution of matter. In a universe with multiple types of matter fluctuations, as is the case with massive neutrinos, the relation between the halo field and the matter fluctuations may be more complicated. We develop a method for calculating the linear bias factor relating fluctuations in the halo number density to fluctuations in the mass density in the presence of multiple fluctuating components of the energy density. In the presence of massive neutrinos we find a small but pronounced feature in the halo bias near the neutrino free-streaming scale. The neutrino feature is a small step with amplitude that increases with halo mass and neutrino mass density. The scale-dependent halo bias lessens the suppression of the small-scale halo power spectrum and should therefore weaken constraints on neutrino mass from the galaxy auto-power spectrum and correlation function. On the other hand, the feature in the bias is itself a novel signature of massive neutrinos that can be studied independently.  
\end{abstract}
\maketitle
\section{Introduction}
Mapping the large-scale structure of the universe, e.g. through the large-scale distribution of galaxies and quasars, is a primary means for learning about cosmology. The clustering statistics of cold dark matter (CDM) halos hosting galaxies and quasars contains key information about the late-time expansion history and matter contents -- in particular the energy density in massive neutrinos \cite{Bond:1980ha, Hu:1997mj, Seljak:2006bg,Saito:2009ah, Reid:2009nq, Thomas:2009ae, Swanson:2010sk, Xia:2012na, RiemerSorensen:2011fe,Zhao:2012xw,dePutter:2012sh}. A crucial ingredient in interpreting measurements of the galaxy clustering is an understanding of how the large-scale halo distribution traces the large-scale distribution of mass.  On very large scales, there is a linear relationship between the fluctuations in the matter density $\delta_m \equiv \delta\rho_m/\rho_m$ and fluctuations in the number density of halos $\delta_n \equiv \delta n/n$
\be
\label{eq:bdef}
\delta_n \approx b\, \delta_m
\ee
where $b$ is the halo bias. The matter density $\rho_m$ consists of CDM $\rho_c$, baryons $\rho_b$, and a tiny fraction of massive neutrinos $\rho_\nu$. 

On scales larger than the baryonic Jeans scale, the behavior of CDM and baryons is indistinguishable. For the purposes of calculating the gravitational evolution of structure on these large scales, CDM and baryons may be treated as a single fluid. (Indeed, for the rest of this paper we treat them as a single fluid identified by the subscript $c$.) Cosmic background neutrinos have a temperature $T_\nu \approx 1.95K$ and therefore neutrinos of mass $m_\nu$ may have a large thermal velocity $u_\nu \sim m_\nu/T_\nu$ that permits them to free-stream out of overdense regions. Perturbations in the neutrino energy density therefore differ from perturbations in the CDM and baryons on scales smaller than the neutrino free-streaming length $\lambda_{fs} \sim u_\nu/(aH)$ where $a$ is the scale factor and $H$ is the Hubble parameter (for a review of cosmology with massive neutrinos see \cite{Lesgourgues:2006nd}).  The absence of neutrino perturbations on small scales reduces the amplitude of $\delta_m$ and further slows the growth of small-scale perturbations in CDM. On the largest scales perturbations in the neutrino energy density and CDM behave indistinguishably.  Massive neutrinos, therefore, cause the evolution of density perturbations $\delta_m$ and $\delta_c$ to be scale dependent. 

The absolute value of the neutrino mass has yet to be detected. Neutrino oscillation data in combination with the inferred relic abundance of neutrinos \cite{Ade:2013zuv, Hinshaw:2012aka, Hou:2012xq, Sievers:2013ica} require that massive neutrinos contribute at least a few tenths of a percent to the cosmic energy budget today $\Omega_\nu h^2 \gsim 0.06\,eV/94\,eV$ \cite{Beringer:1900zz}. Current bounds on the sum of the neutrino masses from cosmological datasets are $\sum_i m_{\nu i} \lsim 0.2 \,eV - 1\,eV$, depending on the dataset (see e.g. \cite{Ade:2013zuv, Hinshaw:2012aka, Hou:2012xq, Sievers:2013ica} for cosmic microwave background constraints, \cite{Seljak:2006bg, Reid:2009nq, Thomas:2009ae, Swanson:2010sk, Saito2010, Xia:2012na, RiemerSorensen:2011fe,Zhao:2012xw,dePutter:2012sh,Beutler:2013yhm} for constraints from galaxy and Lyman-alpha forest surveys, and \cite{Vikhlinin:2008ym,Mantz:2009rj,Benson:2011uta,Reichardt:2012yj,Hasselfield:2013wf,Ade:2013lmv} for constraints from the abundance of galaxy clusters). 

The purpose of this paper is to develop an analytic model for halo bias in the presence of massive neutrinos. As we shall see massive neutrinos generate a scale-dependent feature in the halo bias near the neutrino free-streaming scale. The scale dependence of the bias arises from two effects: (i) the scale-dependent growth of density perturbations causes the Lagrangian halo bias with respect to the CDM to be scale dependent and (ii) the halo field traces the CDM density fluctuations, rather than the total mass fluctuations, and the scale-dependent relationship between $\delta_c$ and $\delta_m = f_c\delta_c + f_\nu\delta_\nu$ causes additional scale dependence in the relationship between $\delta_n$ and $\delta_m$. 

The authors of  \cite{Hui:2007zh, Parfrey:2010uy} noted that scale-dependent growth gives rise to scale-dependent halo bias and studied this in detail for cosmologies with scale-dependent growth associated with the late-time accelerated expansion. In the case of massive neutrinos, the scale-dependent growth starts at earlier times and this motivates us to develop a framework for calculating halo bias in a cosmology with perturbations in multiple components of the energy density at early times. Our approach is simply to solve the spherical collapse model for halo abundance \cite{Gunn:1972sv} in the presence of long-wavelength fluctuations in the energy density and pressure of all of the different constituents. The spherical collapse results can then be used as input to the peak-background split calculation of the halo bias \cite{Sheth:1999mn,Cole:1989vx}. The spherical collapse model for halo abundance is, at best, a crude approximation to halo formation. Nevertheless, the halo bias factor calculated from the spherical collapse model and the peak-background split is relatively robust \cite{Manera:2009ak}. Our goal here is simply to estimate the amplitude and develop a physical understanding of the effect of massive neutrinos on halo bias and for this purpose the spherical collapse model is sufficient. The analysis here assumes that neutrino clustering interior to halos is unimportant for calculating halo evolution and for identifying the mass of the halo. That is the neutrino contribution to the halo mass is negligible. This should be a safe assumption for the range of neutrino masses we consider \cite{LoVerde:2013lta, LoVerde:2014rxa}. 

There are of course an increasing number of N-body simulations of large-scale structure in neutrino + CDM cosmologies (see e.g. \cite{Colin:2007bk,Brandbyge:2008rv,Agarwal:2010mt, Viel:2010bn,Viel:2010bn,Brandbyge:2010ge,Marulli:2011he,VillaescusaNavarro:2012ag,Upadhye:2013ndm} and  \cite{Colin:2007bk,Brandbyge:2008rv,Viel:2010bn,Brandbyge:2010ge,VillaescusaNavarro:2012ag, Villaescusa-Navarro:2013pva, Castorina:2013wga,Costanzi:2013bha} for simulations that include both neutrino and CDM particles). Interestingly, the simulations of \cite{Villaescusa-Navarro:2013pva, Castorina:2013wga} (which appeared while this work was in preparation) show evidence for the scale-dependent halo bias described here. Where possible we make a comparison between our calculations and those results. It would be very exciting to make a systematic comparison between halo bias from simulations and our predictions. 

The effect discussed here is a neutrino-induced scale-dependent correction to the very large-scale ($k  < 0.1 \,Mpc^{-1}$) linear bias. Even in CDM-only cosmologies the constant linear bias model is too simplistic. In recent years there has been substantial progress in developing more sophisticated models of halo biasing (see, e.g. \cite{Smith:2006ne, Matsubara:2008wx, McDonald:2008, Desjacques2010, Carlson:2012, Sheth:2012, Chan2012, Baldauf:2013hka,Tassev:2013zua, Saito:2014qha, Biagetti:2014pha} and references therein). Some of the additional ingredients (e.g. including nonlinear gravitational evolution, nonlinear halo biasing, imposing the constraint that proto-halos live in peaks of the initial density field, and halo exclusion effects) introduce additional sources of scale dependence to the relationship between the statistics of the halo field and the dark matter field. These contributions to scale dependence are primarily important on smaller scales ($k\gsim 0.1 Mpc^{-1}$, which is generally smaller than the neutrino free-streaming scale) so we do not include them here and instead truncate our predictions for the halo bias at $k\sim 0.1 Mpc^{-1}$. Such additional ingredients will be necessary to model the galaxy power spectrum and galaxy-matter cross-power spectrum across the entire observable range of scales but a complete model is beyond the scope of this paper. 

Reference \cite{Biagetti:2014pha}, which appeared after this paper was completed, finds scale-dependent halo bias that is changed in the presence of massive neutrinos. They find that massive neutrinos alter the amplitude of the coefficients of terms that are nonlinear in the density field, the coefficient of a $k^2$-term that appears from the peak constraint (corrections not considered in this paper), and change the amplitude of the bias above the neutrino free-streaming scale because halos trace fluctuations in the CDM density rather than total matter density (one of two contributions to the bias feature identified in this paper and also discussed in \cite{Villaescusa-Navarro:2013pva, Castorina:2013wga}). The additional large-scale, scale-dependent feature in this paper (the scale-dependent Lagrangian bias with respect to CDM) is only $\lsim (few)\%$ for $\sum_i m_{\nu i}\le 0.6\, eV$ so there does not appear to be any contradiction with \cite{Biagetti:2014pha}, which finds agreement between their prediction and the simulations of \cite{Villaescusa-Navarro:2013pva} at the $\sim 3\%$ level for the same neutrino mass range. 

The rest of this paper is organized as follows. In \S \ref{sec:halobiasoutline} we outline the calculation of halo bias in a cosmology with perturbations in additional non-CDM components. In \S \ref{sec:nuLCDM} we review the calculation of spherical collapse in a $\nu \Lambda CDM$ universe, and then in \S \ref{sec:SCLW} develop the calculation of spherical collapse in the presence of long-wavelength fluctuations in the energy density. Numerical results for the spherical collapse threshold in the presence of long-wavelength modes are presented in \S \ref{sec:numericalSC}. In \S \ref{sec:numericalbias} we combine the results from \S \ref{sec:halobiasoutline} and \S\ref{sec:SCLW} to calculate the scale-dependent halo bias. Conclusions and a discussion of future directions are given in \S \ref{sec:conclusions}. 

\section{Halo bias in a mixed dark matter cosmology}
\label{sec:halobiasoutline}
We calculate the halo power spectrum in a mixed dark matter (neutrino + CDM) universe. In a universe with neutrino and CDM perturbations, the fluctuations in the total matter density are
\ba
\delta_m &=& \frac{\delta\rho_c + \delta\rho_\nu}{\rho_c+ \rho_\nu}\,,\\
& \equiv& f_c\delta_c+ f_\nu\delta_\nu\,,
\ea
where $f_c$, $f_\nu$ are the fractions of the matter density that are cold dark matter and neutrinos ($f_c +f_\nu = 1$)  and $\delta_c$, $\delta_\nu$ are the fractional perturbations to the CDM and neutrino energy densities. In the standard cosmology, the neutrino and CDM perturbations are coherent on large scales ($k \ll k_{fs}$ where $k_{fs}$ is the neutrino free-streaming scale), but on scales below the neutrino free-streaming scale, neutrino perturbations are damped. 

\begin{figure}[t]
\begin{center}
$\begin{array}{cc}
 \includegraphics[width=0.5\textwidth]{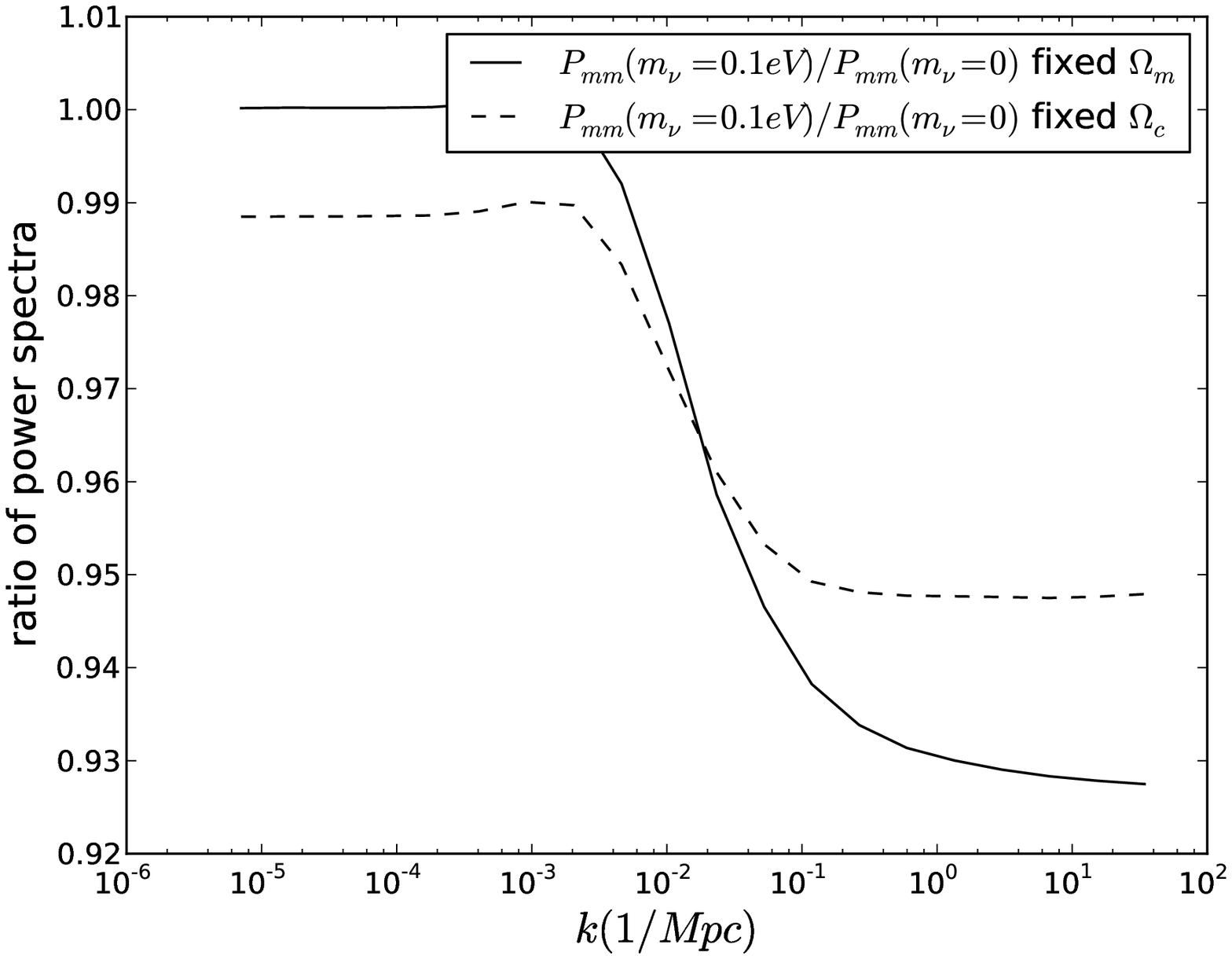} & \includegraphics[width=0.5\textwidth]{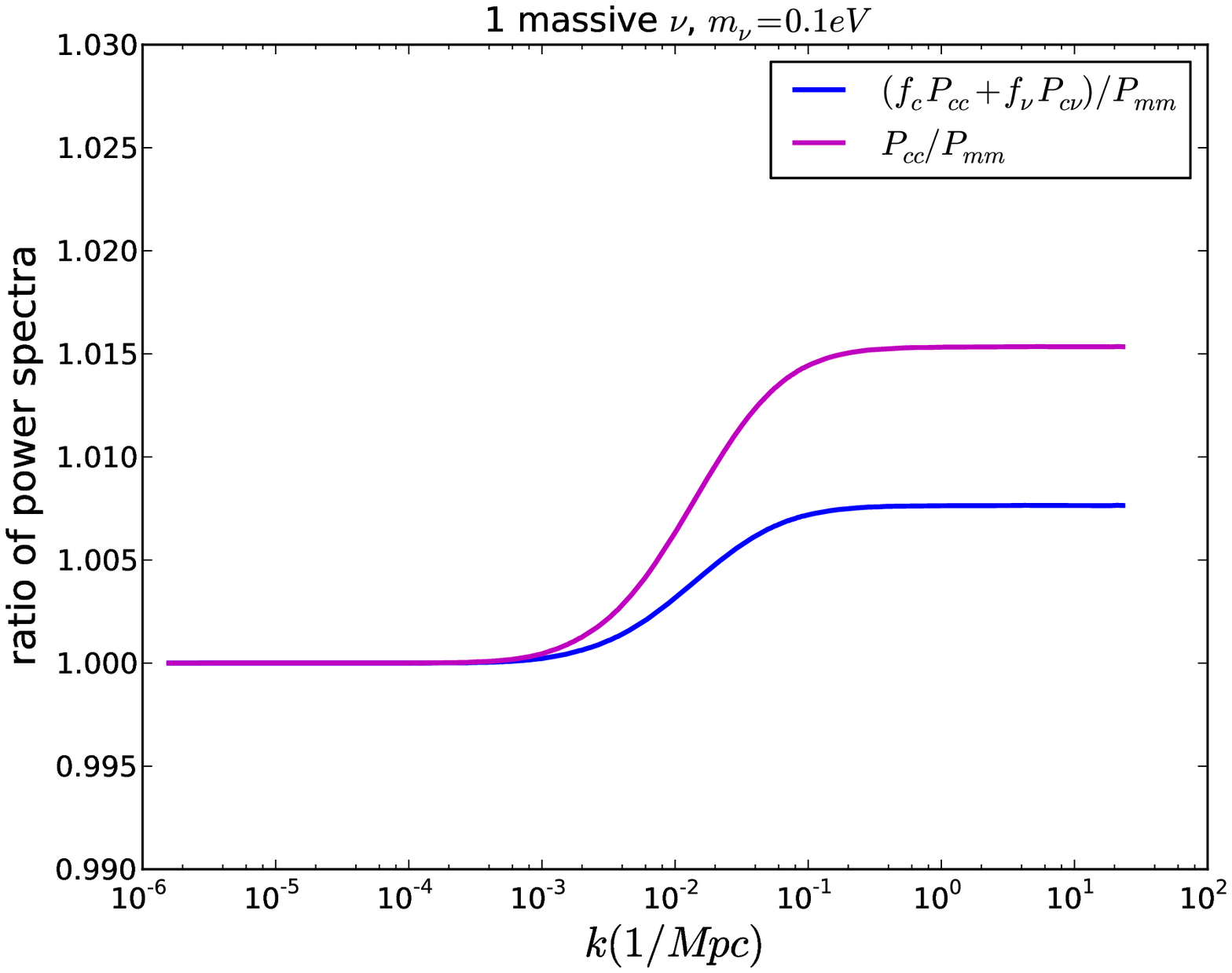}
\end{array}$
\caption{\label{fig:powerratios} Left: The scale-dependent changes to the matter power spectrum in a cosmology with a single massive neutrino $m_{\nu} = 0.1\,eV$. Right: The ratio of the CDM power spectrum and CDM-matter cross-power spectrum to the matter power spectrum in a cosmology with a single massive neutrino $m_{\nu} = 0.1\,eV$. Both quantities are plotted at $z =0$.}
\end{center}
\end{figure}

Long-wavelength fluctuations in the different matter components can modulate the number density of dark matter halos. But the way that long-wavelength perturbations in CDM alter halo abundance may be different from the way long-wavelength perturbations in the neutrino density alter halo abundance. For now, we write fluctuations in the number density of halos as
\be
\label{eq:bx}
\delta_n = \sum_X b_X\delta_{X,L}\,,
\ee
where $\delta_{X,L}$ is a long-wavelength fluctuation in CDM, baryons, neutrinos or whichever component of the energy density we are considering and $b_X$ are the to-be-determined bias factors. For adiabatic initial conditions, a single parameter specifies the amplitude of fluctuations in each component. For instance, we can specify the amplitude of the long-wavelength CDM perturbation, $\delta_{c}(k_L,z)$ and the perturbations in other components are given by
\be
\delta_{X}(k,z) = \frac{T_{X}(k,z)}{T_{c}(k,z)}\delta_{c}(k,z)\,,
\ee
where $T_{c}(k)$ is the transfer function for cold dark matter and $T_X(k,z)$ is the transfer function for component $X$ (i.e. $X = c$, $\nu$, and $\gamma$). 

To determine the bias factors in Eq.~(\ref{eq:bx}), we adopt the peak-background split argument \cite{White1987, Bardeen:1985tr, Cole:1989vx}, namely, that the critical value of the halo-scale CDM density perturbation required for a halo to form is modulated by the background density.  We calculate the critical overdensity using the spherical collapse model in the presence of a long-wavelength, adiabatic fluctuation in all the matter components.  From this, we can determine change in the value of the critical amplitude density fluctuation in CDM required for a spherical halo to collapse by redshift $z$, 
\be
\label{eq:ddeltacrit}
\frac{d\delta_{crit}}{d\delta_{c,L}}(k) = \frac{\delta_{crit}(z|\delta_{c,L}(k)) - \delta_{crit}(z | \delta_{c,L} = 0)}{\delta_{c,L}(k)}\,,
\ee
and we have allowed for the possibility that the derivative above depends on the wavenumber of the long-wavelength mode $k$. Again, for adiabatic initial conditions specifying the amplitude of the initial fluctuation in one component is sufficient to determine the linear fluctuations in all other components and we choose to use the amplitude of the CDM and baryon fluctuation $\delta_{c,L}$ to specify the long wavelength modes. 

Equation~(\ref{eq:ddeltacrit}) along with an expression for the halo mass function $n$ gives the Lagrangian bias as 
\be
\label{eq:bXLagrangian}
b_X^{Lagrangian} \equiv \frac{\partial \ln n}{\partial \delta_{crit}}\frac{d\delta_{crit}}{d\delta_{X,L}}\,.
\ee
To map between the Lagrangian halo density and the final, Eulerian halo density we need to relate the volumes in Lagrangian and Eulerian space.  The cold dark matter mass in an infinitesimal volume $d^3\x^L$ is conserved (unlike the neutrino mass, which will stream out of small regions due to the large peculiar velocities). Using the CDM mass to label volumes gives the relationship between Eulerian and Lagrangian volume elements as
\be
\label{eq:EulerianLagrangian}
(1 + \delta_c^{Eulerian})d^3\x^E = (1+ \delta_c^{Lagrangian})d^3 \x^L \approx d^3 \x^L\,.
\ee
Note that if neutrinos, or some other component were tracking CDM identically, we could use any one of them to map between Eulerian and Lagrangian volumes and our result would not differ.  

Equations~(\ref{eq:bXLagrangian}) and (\ref{eq:EulerianLagrangian}) give a final expression for linear fluctuations in Eulerian number density of halos,
\be
\delta_n =\left(1 + \frac{\partial \ln n}{\partial \delta_{crit}}\frac{d\delta_{crit}}{d\delta_{c,L}}(k)\right)\delta_{c,L}(k)\,.
\ee
With this, the halo-matter cross-power spectrum is given by
\be
P_{nm}(k) = \left(1 + \frac{\partial \ln n}{\partial \delta_{crit}}\frac{d\delta_{crit}}{d\delta_{c,L}}(k)\right)\left(f_cP_{cc}(k) + f_\nu P_{c\nu}(k)\right)
\ee
where $P_{cc}(k)$ is the CDM autopower spectrum, and $P_{c\nu}$ is the CDM-neutrino cross-power spectrum. The halo-halo autopower spectrum is given by
\be
P_{nn}(k) = \left(1 + \frac{\partial \ln n}{\partial \delta_{crit}}\frac{d\delta_{crit}}{d\delta_{c,L}}(k)\right)^2P_{cc}(k)
\ee
and the matter-matter auto-power spectrum is given by
\be
P_{mm}(k) = f_c^2P_{cc}(k) + 2f_cf_\nu P_{c\nu}(k) + f_\nu^2 P_{\nu\nu}(k)\,.
\ee
The observed bias factor is then
\ba
b(k) &\equiv& \frac{P_{nm}(k)}{P_{mm}(k)}\,,\\
\label{eq:bkn}
&=&  \left(1 + \frac{\partial \ln n}{\partial \delta_{crit}}\frac{d\delta_{crit}}{d\delta_{c,L}}(k)\right)\frac{f_cP_{cc}(k) + f_\nu P_{c\nu}(k)}{f_c^2P_{cc}(k) + 2f_cf_\nu P_{c\nu}(k) + f_\nu^2 P_{\nu\nu}(k)}\,.
\ea

The neutrino-induced suppression in the matter power spectrum, along with the ratio of the CDM-matter cross-power spectrum to the matter-matter auto-power spectrum needed in Eq.~(\ref{eq:bkn}) are plotted in Fig.~\ref{fig:powerratios}. On scales that are large compared to the neutrino free-streaming scale, $P_{cc} \approx P_{c\nu} \approx P_{\nu\nu}$, while on smaller scales $P_{\nu\nu}\approx P_{c\nu} \approx 0$. This results in 
\be
\label{eq:bkestimate}
b(k) = \left(1 + \frac{\partial \ln n}{\partial \delta_{crit}}\frac{d\delta_{crit}}{d\delta_{c,L}}(k)\right) \left\{\begin{array}{cc} 1 & k \ll k_{fs} \\ 1/f_c & k\gg k_{fs} \end{array}\right\} \,,
\ee
where $k_{fs}$ is the comoving neutrino free-streaming scale at redshift $z$ defined by
\be
\label{eq:kfs}
k_{fs} = \frac{\sqrt{3/2}m_{\nu} H(z)}{3.15T_\nu(1+z)}\,.
\ee
Equation ~(\ref{eq:bkn}) is our final expression for the scale-dependent halo bias. In the next section we outline the calculation of $d\delta_{crit}/d\delta_{c,L}(k)$ in the spherical collapse model which we can then use in Eq.~(\ref{eq:bkn}) to calculate $b(k)$. 

The scale dependence in Eqs.~(\ref{eq:bkn})-(\ref{eq:bkestimate}) is, of course, not the only change to the halo bias from massive neutrinos. The suppression in the CDM power spectrum suppresses the variance of mass fluctuations on scale $M$,
\be
\label{eq:sigmaM}
\sigma^2(M,z) = \int_0^{\infty} \frac{k^2 dk}{2\pi^2} |W(kR(M))|^2 P_{cc}(k,z)
\ee
where $W(kR) = 3\sin(kR)/(kR)^3 - 3\cos(kR)/(kR)^2$ and $R = (3M/(4\pi\bar\rho_c))^{1/3}$. At fixed halo mass, the decrease in $\sigma(M,z)$ decreases the abundance of halos and increases the bias. This change to the halo bias is constant with $k$ and therefore unobservable in measurements of galaxy clustering that treat the overall amplitude as a free parameter.

\section{Review of spherical collapse in $\nu \Lambda CDM$ universe}
\label{sec:nuLCDM}
In $\nu\Lambda CDM$, the scale factor $a$ evolves according to
\be
\label{eq:Friedmann}
H^2(a)=\frac{8\pi G}{3}\left(\bar{\rho}_c(a)+\bar{\rho}_\nu(a)+\bar{\rho}_\gamma(a)+\bar{\rho}_\Lambda\right)\,,
\ee 
where $\rho_c$ is the CDM density,  $\rho_\nu$ the neutrino energy density, $\rho_\gamma$ the photon energy density, and $\rho_\Lambda$ the energy density in cosmological constant.  The energy density and pressure of the neutrinos is given by
\be
\bar{\rho}_\nu = 2 \sum_i \int \frac{d^3 p}{(2\pi)^3}\frac{\sqrt{p^2 + m_{\nu i}^2}}{e^{p/T_\nu}+1}\,,\quad\bar{P}_\nu = 2\sum_i  \int \frac{d^3 p}{(2\pi)^3}\frac{p^2}{3\sqrt{p^2 + m_{\nu i}^2}}\frac{1}{e^{p/T_\nu}+1}\,,
\ee
where the neutrino temperature is given by $T_\nu(a)  = 1.95491 K/a$. The other components evolve as $\rho_c\propto 1/a^3$, $\rho_\gamma \propto 1/a^4$, and $\rho_\Lambda = const.$.

\subsection{Equation of motion for $R$}
\label{ssec:Reom}
The sub-horizon equation of motion for a spherical mass shell of radius $R$ enclosing constant (CDM + baryon) mass $M$ is 
\be
\label{eq:ddotRall}
\ddot{R}=-\frac{GM}{R^2}-\frac{4\pi G \int_0^{R} dr r^2 (\rho_{rest}(r,t)+3P_{rest}(r,t))}{R^2}\,,
\ee
where $\rho_{rest}$ and $P_{rest}$ are the energy density and pressure of radiation, neutrinos, and cosmological constant.  The condition $M = \frac{4}{3}\pi R^3\bar{\rho}_{c}(1+\delta_{c})$ relates $R$ to $\delta$ and allows us to set the initial conditions for $R$ in terms of $\delta_{c,i}$ and $\dot \delta_{c,i}$
\be
\label{eq:Rics}
R_i = \bar{R}_i\left(1 - \frac{1}{3}\delta_{c,i}\right) \,, \qquad \dot{R_i} = H_i\bar{R}_i\left(1 - \frac{1}{3}\delta_{c,i} - \frac{1}{3}H_i^{-1}\dot{\delta}_{c,i}\right) \,,\qquad \bar{R}_i =  \left(\frac{3 M}{4\pi\bar{\rho}_{c}}\right)^{1/3}\,.
\ee
The CAMB code can be used to find the numerical value of $\dot\delta_{c, i}/\delta_{c}$ at any redshift for adiabatic initial perturbations \cite{Lewis:1999bs, LoVerde:2014rxa}. 

The final expression that we use to solve for the subhorizon, non-linear evolution of $R(t)$ is then, 
\be
\label{eq:ddotR}
\ddot{R}=-\frac{GM}{R^2}-\frac{4\pi G}{3} \left(2\bar{\rho}_\gamma(t)+\bar{\rho}_\nu(t) + 3\bar{P}_\nu-2\bar{\rho}_\Lambda(t)\right)R\,,
\ee
where $\rho_\gamma$ is the photon energy density, $\rho_\nu$ and $P_\nu$ the energy density and pressure of neutrinos, and $\rho_\Lambda$ the energy density in cosmological constant. In Eq.~(\ref{eq:ddotR}) we have ignored any terms due to  gravitational clustering of neutrinos (or anything other than CDM and baryons) because they are small for the range of neutrino masses we consider \cite{LoVerde:2013lta,LoVerde:2014rxa}. 

\section{Spherical collapse on a long-wavelength mode}
\label{sec:SCLW}
We now consider spherical collapse in the presence of a longer-wavelength density perturbation, which may include CDM, baryons, neutrinos, or photons. The equation of motion in the presence of a long-wavelength mode is

\be
\label{eq:ddotRLW}
\ddot{R}=-\frac{GM}{R^2}-\frac{4\pi G}{3} \left(2{\rho}_\gamma(t) + 2{\rho}_{\nu{\rm{massless}}}(t) + \rho_{\nu{\rm massive}}(t)+3\bar{P}_{\nu{\rm massive}}(t)-2\bar{\rho}_\Lambda(t)\right)R \,.
\ee
The energy densities above (without the $\bar{\quad}$) are given by 
\ba
\rho_\gamma &=& \bar{\rho}_\gamma(1 + \delta_{\gamma,L}(t))\,, \\
\rho_{\nu {\rm {\tiny massless}}} &=& \bar{\rho}_{\nu{\rm {\tiny massless}}}(1 + \delta_{{\nu {\rm {\tiny massless}}},L}(t))\,, \\
\rho_{\nu {\rm {\tiny massive}}} &=& \bar{\rho}_{\nu{\rm {\tiny massive}}}(1 + \delta_{{\nu {\rm {\tiny massive}}},L}(t)) \,.
\ea
The small-scale density fluctuation $\delta_{c,S}$ is defined relative to the local background density which includes the large-scale fluctuation, i.e. $\delta_{c,S} = \rho_c/(\bar{\rho}_c(1+\delta_{c,L})) - 1$. The long-wavelength perturbation in the CDM and baryon density, $\delta_{c,L}$ does not appear in the equation of motion for $R$, but appears in the expression relating the halo mass to the radius  $M = \frac{4}{3}\pi R^3\bar{\rho}_{c}(1+ \delta_{c,S})(1+\delta_{c,L})$. The initial conditions are then
\be
\label{eq:RicsLW}
R_i = \bar{R}_i\left(1 - \frac{1}{3}\left(\delta_{c,iS} + \delta_{c,iL}\right) \right) \,, \qquad \dot{R_i} = H_iR_i\left(1  - \frac{1}{3}H_i^{-1}\left(\dot{\delta}_{c,iS}+\dot{\delta}_{c,iL}\right)\right) \,,\qquad \bar{R}_i =  \left(\frac{3 M}{4\pi\bar{\rho}_{c}}\right)^{1/3}\,.
\ee
We set the initial velocity of the small-scale perturbations by $\dot{\delta}_{c,iS} = \dot{\sigma}(M)/\sigma(M)\delta_{c, iS}$  and will linearly extrapolate $\delta_{c,iS}$ to the collapse time using $\delta_{c,S}(z) = \sigma(M,z)/\sigma(M,z_i)\delta_{c,iS}$ as in \cite{LoVerde:2014rxa}. However, we have checked that for all neutrino and halo masses considered in this paper the amplitude of the neutrino feature in the halo bias is unchanged if we instead used the initial velocity and linear evolution for an exactly top-hat small-scale perturbation $\delta_{c,iS}$.

For a fixed amplitude perturbation in CDM and baryons at $z_i$ given by $\delta_{c,i}$, the corresponding perturbations in the other components are given by
\ba
\label{eq:deltagammaL}
\delta_{\gamma}(\k,z) &=& \delta_{c,i}(\k) \frac{T_{\gamma}(k,z)}{T_{c}(k,z_i)}\,,\\
\label{eq:deltanumlessL}
\delta_{\nu{\rm {\tiny massless}}}(\k,z) &=& \delta_{c,i}(\k) \frac{T_{\nu{\rm {\tiny massless}}}(k,z)}{T_{c}(k,z_i)}\,,\\
\label{eq:deltanumL}
\delta_{\nu{\rm {\tiny massive}}}(\k,z) &=& \delta_{c,i}(\k) \frac{T_{\nu{\rm {\tiny massive}}}(k,z)}{T_{c}(k,z_i)}\,,
\ea
where $T_{\gamma}(k,z)$, $T_{\nu{\rm {\tiny massless}}}(k,z)$, $T_{\nu{\rm {\tiny massive}}}(k,z)$, $T_{c}(k,z)$ are the transfer functions, e.g. $\delta_{\nu}(\k,z) = T_\nu(k,z) \zeta(\k)$ and $\delta_{c}(\k,z) =  T_{c}(k,z)\zeta(\k)$ -- standard output from CAMB.  

We consider spherically symmetric long-wavelength perturbations with contributions from Fourier modes of a fixed wavelength. That is, for wavenumber $k_L$ and initial amplitude $\delta_{c,iL}$, the Fourier-space perturbation is
\be 
\delta_{c,iL}(\k) = \delta_{c,iL} \frac{(2\pi)^3}{4\pi}\frac{\delta_D(|\k| - k_L)}{k_L^2} \quad {\rm so\, that} \quad  \frac{1}{V_R}\int_{V_R} d^3\x\, \delta_{c,iL}(\x) = W(k_LR) \delta_{c,iL} \approx \delta_{c,iL}
\ee
where $\delta_D$ is the Dirac delta function, $W(kR)$ is a top-hat window function, and the last approximation is valid for $k_L R \ll 1$. Equation (\ref{eq:ddotRLW}), along with the initial conditions in (\ref{eq:RicsLW}), and the expressions for $\delta_{X,L}$ in Eqs. (\ref{eq:deltagammaL})  - (\ref{eq:deltanumL}) are the ingredients needed to determine the effect of long-wavelength density perturbations on the small-scale spherical collapse solution. 

\begin{figure}[t]
\begin{center}
$\begin{array}{cc}
 \includegraphics[width=0.5\textwidth]{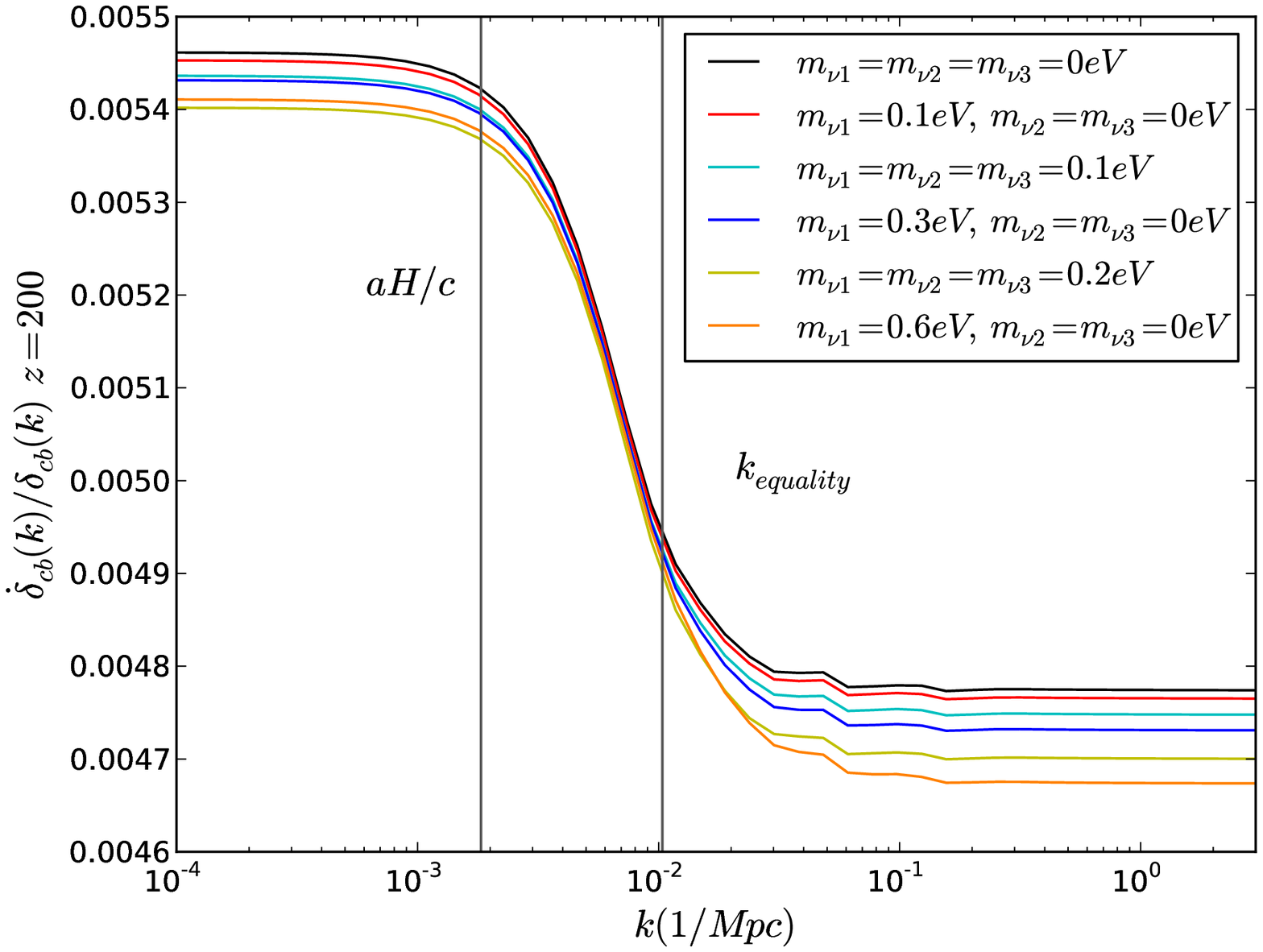} & \includegraphics[width=0.5\textwidth]{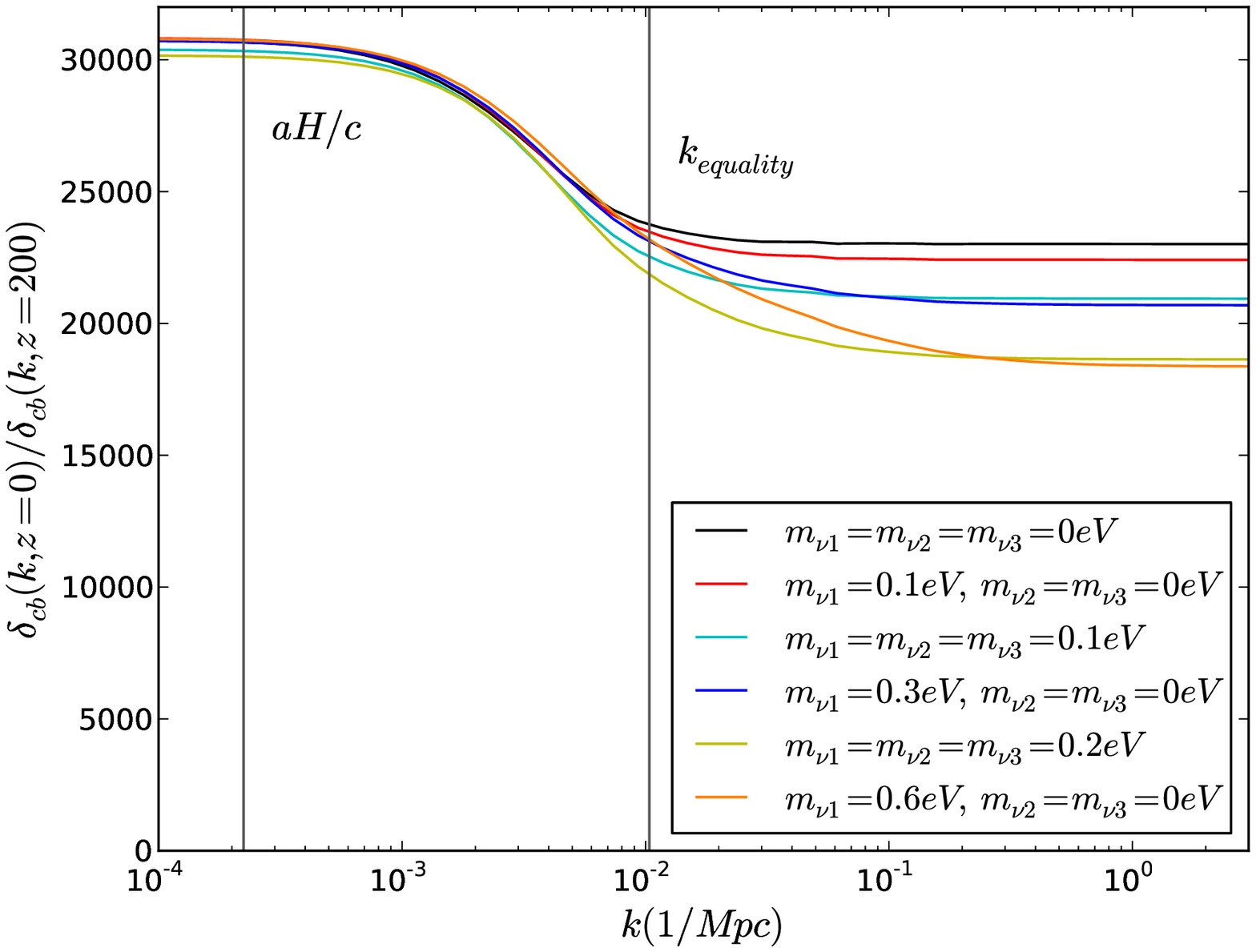}
\end{array}$
\caption{\label{fig:deltaclinears} Left: The scale dependence of $\dot\delta_{c}/\delta_{c}$ at $z_i$, plotted for a number of different neutrino mass hierarchies. Also shown is the horizon scale at $z_i  = 200$ and the matter radiation equality scale. Right: The scale dependence of the linear evolution of CDM and baryon perturbations between $z= 200$ and $z = 0$. In both panels $\Omega_c$ is fixed so varying $\Omega_\nu$ changes the total matter density $\Omega_m$. }
\end{center}
\end{figure}

There are a few points to be made before studying the numerical solutions to Eq.~(\ref{eq:ddotRLW}). First, note that in the absence of $\delta_{X, L}(t)$ and neglecting any difference between $\dot{\delta}_{c,iS}/\delta_{c,iS}$ and $\dot{\delta}_{c,iL}/\delta_{c,iL}$ (as shown in Fig. \ref{fig:deltaclinears}, $\dot\delta_{c,i}/\delta_{c}$ is scale dependent so these terms are different), we have $R(t,\delta_S, \delta_L) = R(t, \delta_S + \delta_L, \delta_L = 0)$. In particular, if $\delta_{c,iS}$ is the critical value of the initial density perturbation for $R$ to have collapsed by $z_{collapse}$ the critical value in a region with a long-wavelength density perturbation $\delta_{c,iL}$ is just shifted to $\delta_{c,Si} - \delta_{c,iL}$.

Furthermore, in this limit the mapping $\delta_{c,iS} \rightarrow \delta_{c,iS} - \delta_{c,iL}$ does not depend on the magnitude or wavelength of $\delta_{c}(\k)$.  Second, we note that even if the relationship $\delta_{c,iS} \rightarrow \delta_{c,iS} - \delta_{c,iL}$ is $k$-independent, when expressed in terms of the linearly evolved quantities $\delta_{c,S}(z)$ and $\delta_{c,L}(\k,z)$ there may be scale dependence if the evolution of $\delta_{c}(\k,z)$ is scale dependent, this is the source of scale-dependent bias in \cite{Parfrey:2010uy}. 

\section{Numerical Results for spherical collapse on a long-wavelength mode}
\label{sec:numericalSC}
In this section we show numerical results for the dependence of the spherical collapse threshold on the amplitude and wavelength of the long-wavelength mode $\delta_{c,L}$, which allows us to calculate the derivative $d\delta_{crit}/d\delta_{c,L}(k)$, and finally the scale-dependent halo bias in Eq.~(\ref{eq:bkn}). 

In the numerical calculations presented here, we solve the equation of motion for $R(t)$ in Eq.~(\ref{eq:ddotRLW}), using the initial conditions in Eq.~(\ref{eq:RicsLW}), and the expressions for $\delta_{\nu,L}$ and $\delta_{\gamma,L}$ in Eqs.~(\ref{eq:deltagammaL})- (\ref{eq:deltanumL}). We use CAMB to calculate all the linear quantities $\delta_{c,iS}$, $\dot\delta_{c,iS}$, $\delta_{c,L}(k,z)$,  $\delta_{\nu,L}(k,z)$, $\delta_{\gamma,L}(k,z)$. We assume a flat $\Lambda CDM$ cosmology with Hubble parameter $h = 0.67$ and baryon density $\Omega_bh^2 =  0.022 $ (we treat baryons and CDM identically). We assume three species of massive neutrinos with variable masses $m_{\nu 1}$, $m_{\nu 2}$ and $m_{\nu 3}$. Massive neutrinos contribute a fraction $\Omega_{\nu }h^2 \approx \sum m_{\nu i}/(94 \,eV)$ to the critical energy density. For fixed CDM and baryon densities, changing the neutrino masses then leads to a different total matter ($\Omega_m = \Omega_c + \Omega_b +  \Omega_\nu$) density today. Plots with this scenario are referred to as with fixed $\Omega_c$. The vacuum energy density $\Omega_\Lambda$ is adjusted to keep the universe flat ($\Omega_\Lambda = 1 - \Omega_c - \Omega_b - \Omega_\nu - \Omega_\gamma$). 
We also study neutrino mass effects at fixed $\Omega_m$ and $\Omega_\Lambda$ by setting $\Omega_c h^2= 0.1199  - \Omega_\nu h^2$. We consider a number of examples of neutrino hierarchies and each figure is labeled with all three masses. In this paper, ``Normal Hierarchy" means $m_{\nu 1} = 0.05\,eV$, $m_{\nu 2} = 0.01\,eV$, $m_{\nu 3}  = 0\,eV$ and ``Inverted Hierarchy" means $m_{\nu 1} = m_{\nu 2} = 0.05\,eV$ and $m_{\nu 3} = 0\,eV$.

\begin{figure}[t]
\begin{center}
$\begin{array}{cc}
 \includegraphics[width=0.5\textwidth]{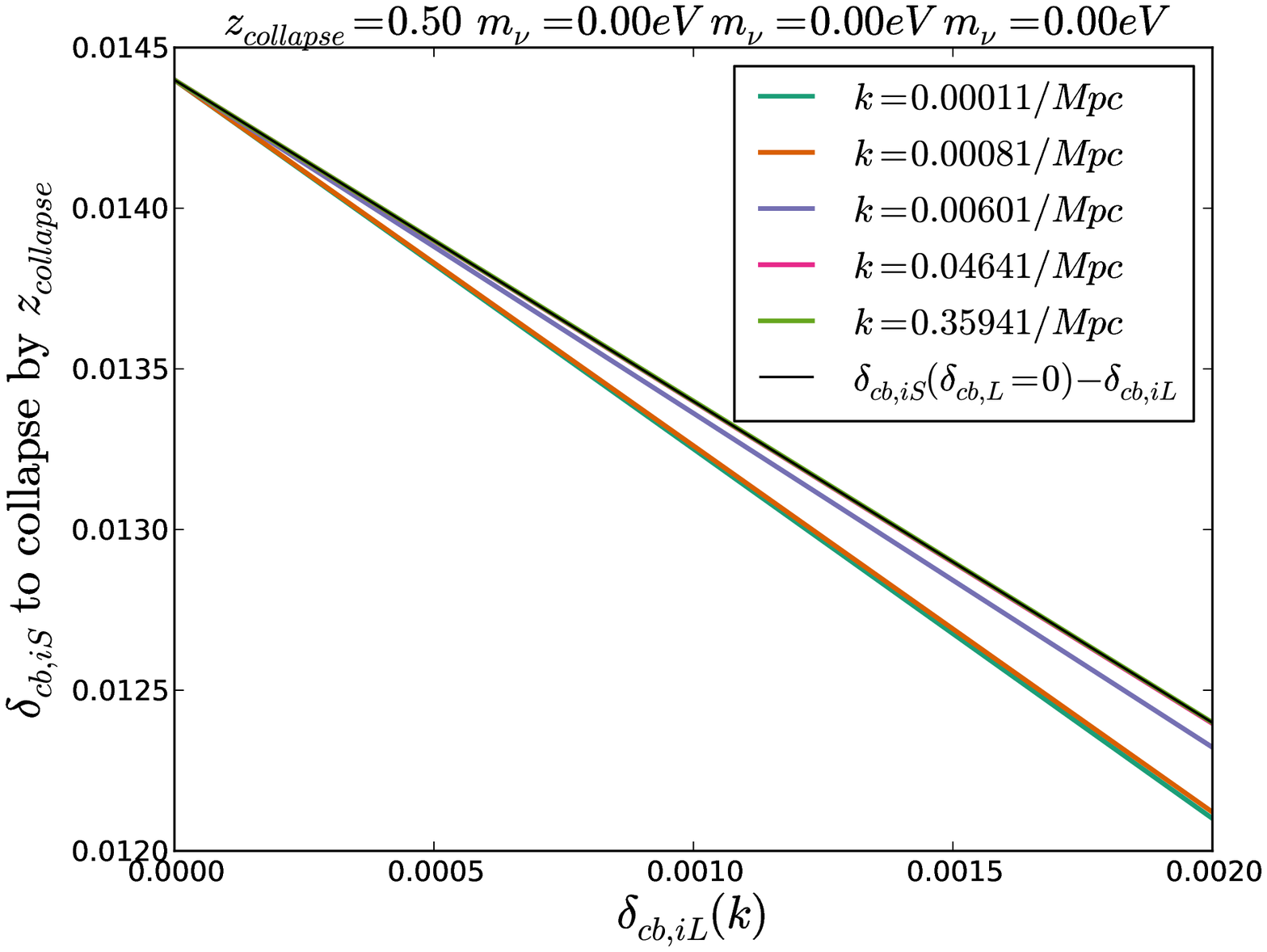} & \includegraphics[width=0.5\textwidth]{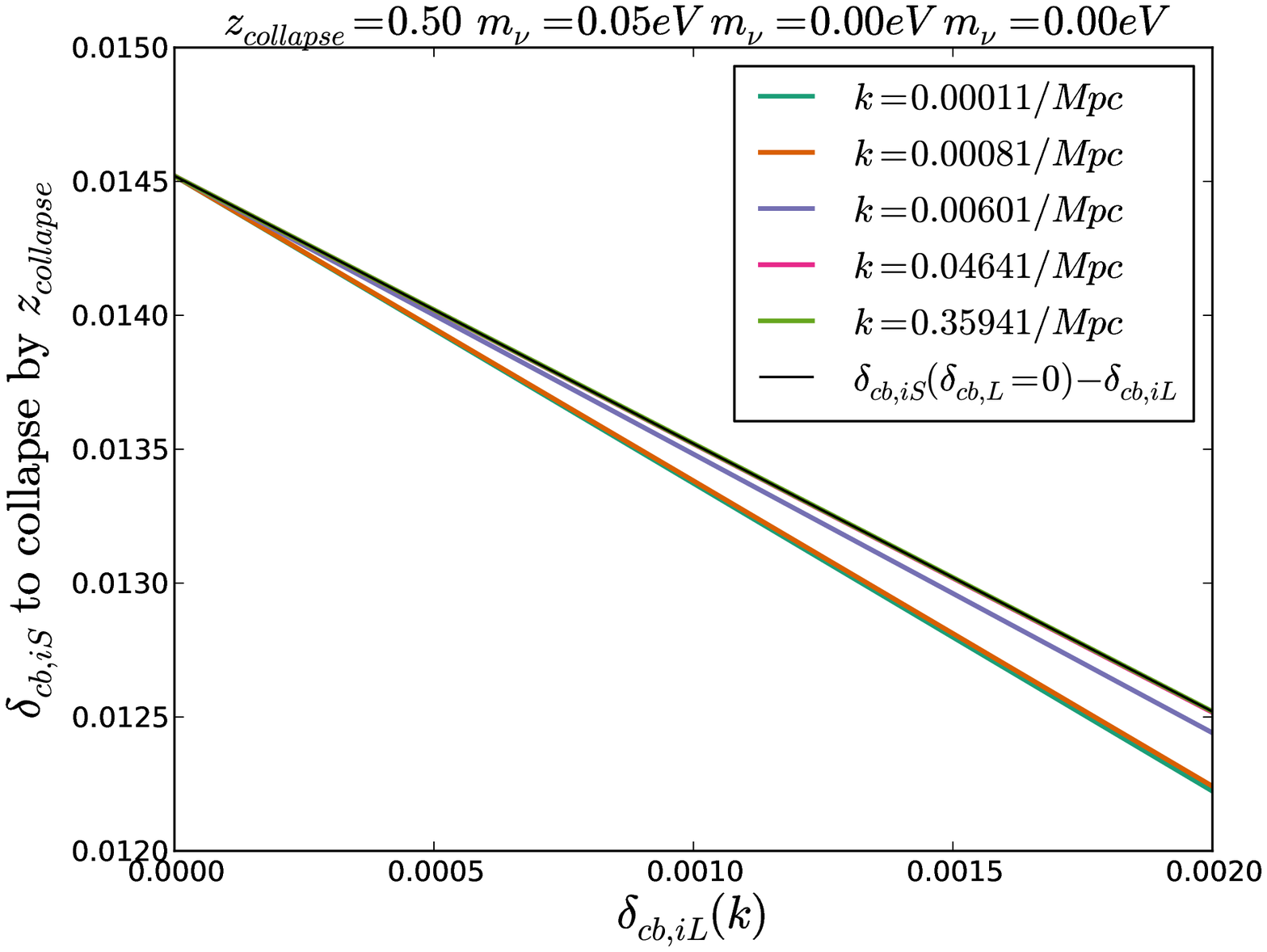}
\end{array}$
\caption{\label{fig:deltaciSdeltaciL} Plotted is the relationship between the initial perturbation values $\delta_{c,iS}$ and $\delta_{c,iL}(k)$ for halos of $M = 10^{13}M_\odot$ that collapse at the same time $z_{collapse} =0.5$ for a range of $k$, the wave number of the long-wavelength mode. Left: $m_{\nu 1} = m_{\nu 2} = m_{\nu 3} = 0\,eV$, Right: $m_{\nu 1} = 0.05\,eV$, $m_{\nu 2} = m_{\nu 3} = 0\,eV$.}
\end{center}
\end{figure}

\begin{figure}[t]
\begin{center}
$\begin{array}{cc}
 \includegraphics[width=0.5\textwidth]{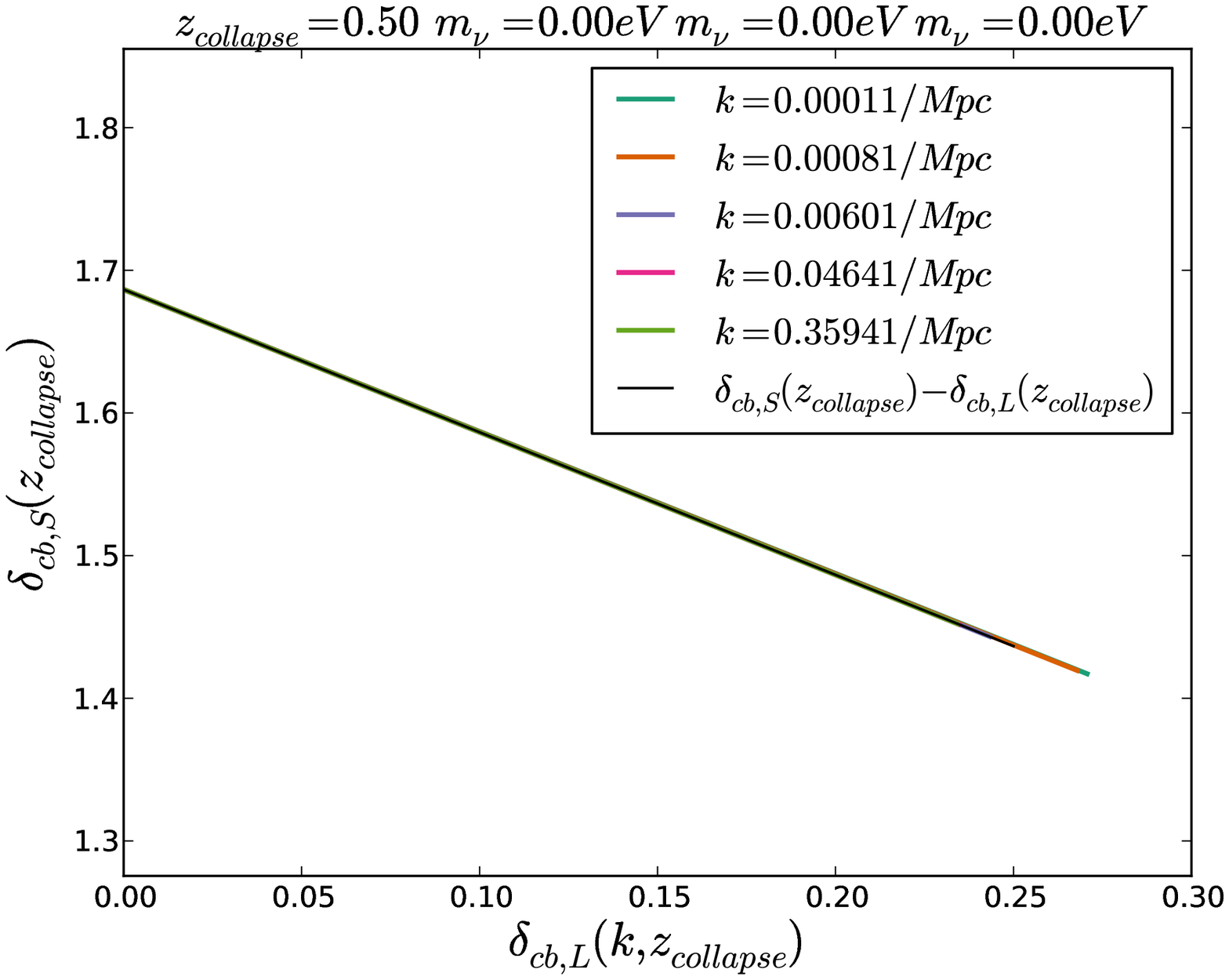} & \includegraphics[width=0.5\textwidth]{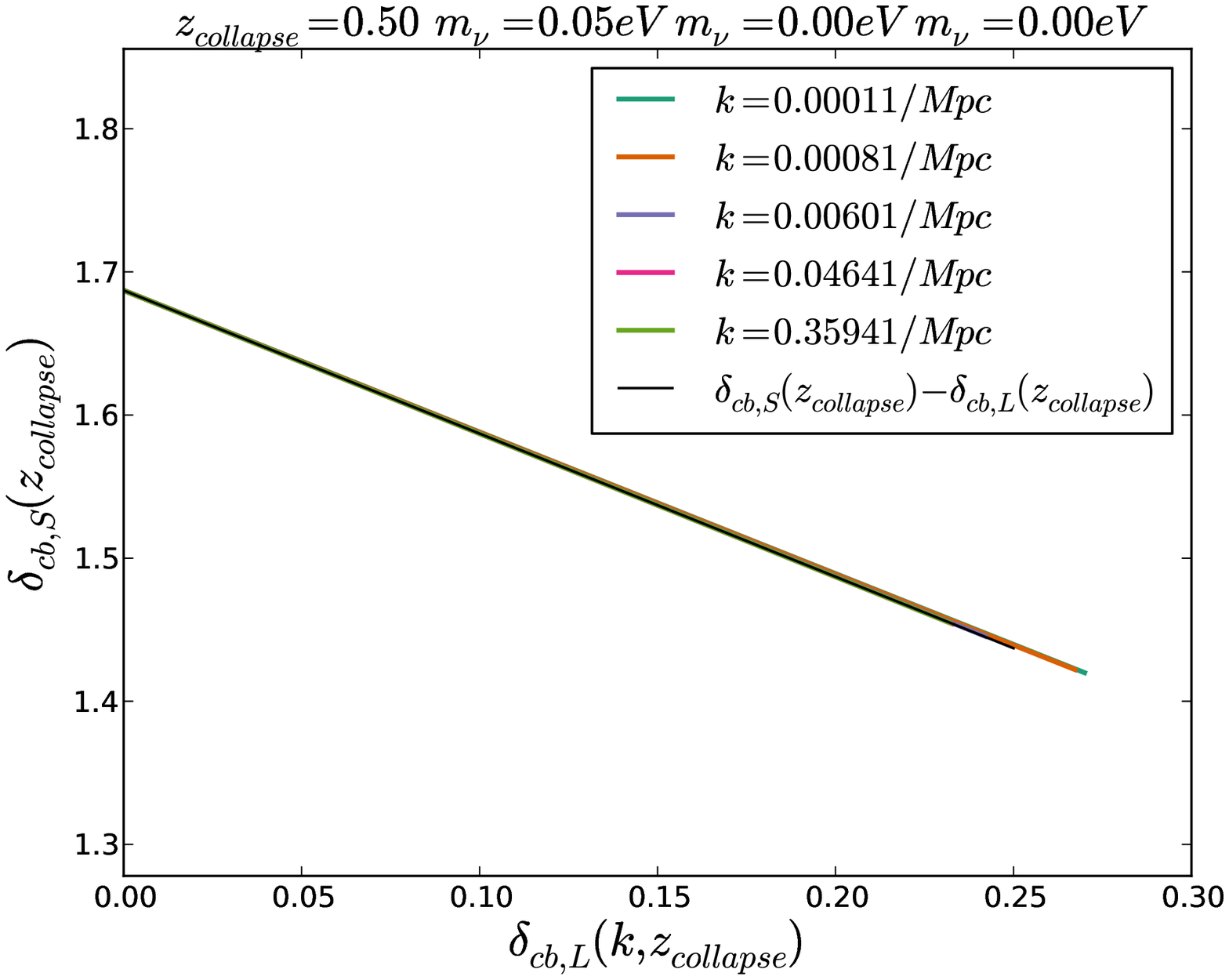}
\end{array}$
\caption{\label{fig:deltaczSdeltaczL} Plotted is the relationship between $\delta_{c,S}(z_{collapse})$ and $\delta_{c,iL}(k,z_{collapse})$ (the initial perturbation amplitudes linearly extrapolated to the collapse redshift) for halos of $M = 10^{13}M_\odot$ that collapse at the same time $z_{collapse} =0.5$ for a range of values of $k$. Left: $m_{\nu 1} = m_{\nu 2} = m_{\nu 3} = 0\,eV$, Right: $m_{\nu 1} = 0.05\,eV$, $m_{\nu 2} = m_{\nu 3} = 0\,eV$.}
\end{center}
\end{figure}

First, we consider the effect of a long-wavelength mode on the critical value of $\delta_{c,iS}$ required to have collapsed by redshift $z_{collapse}$. In Fig. \ref{fig:deltaciSdeltaciL}, we plot this quantity for a cosmology with massless neutrinos only and a cosmology with a single massive neutrino of mass $m_{\nu} = 0.05\,eV$ for a range of values of $k$, the wave number of the long-wavelength mode. In both cases, the relationship between $\delta_{crit, i}$ and $\delta_{c,iL}$ is linear but the slopes vary with the wavelength $k$ of the long-wavelength mode $\delta_{c,L}$. In Fig. \ref{fig:deltaczSdeltaczL} we plot the same quantities linearly extrapolated to $z_{collapse}$: the relationship remains linear but the dependence on $k$ is reduced. 

\begin{figure}[t]
\begin{center}
$\begin{array}{cc}
 \includegraphics[width=0.5\textwidth]{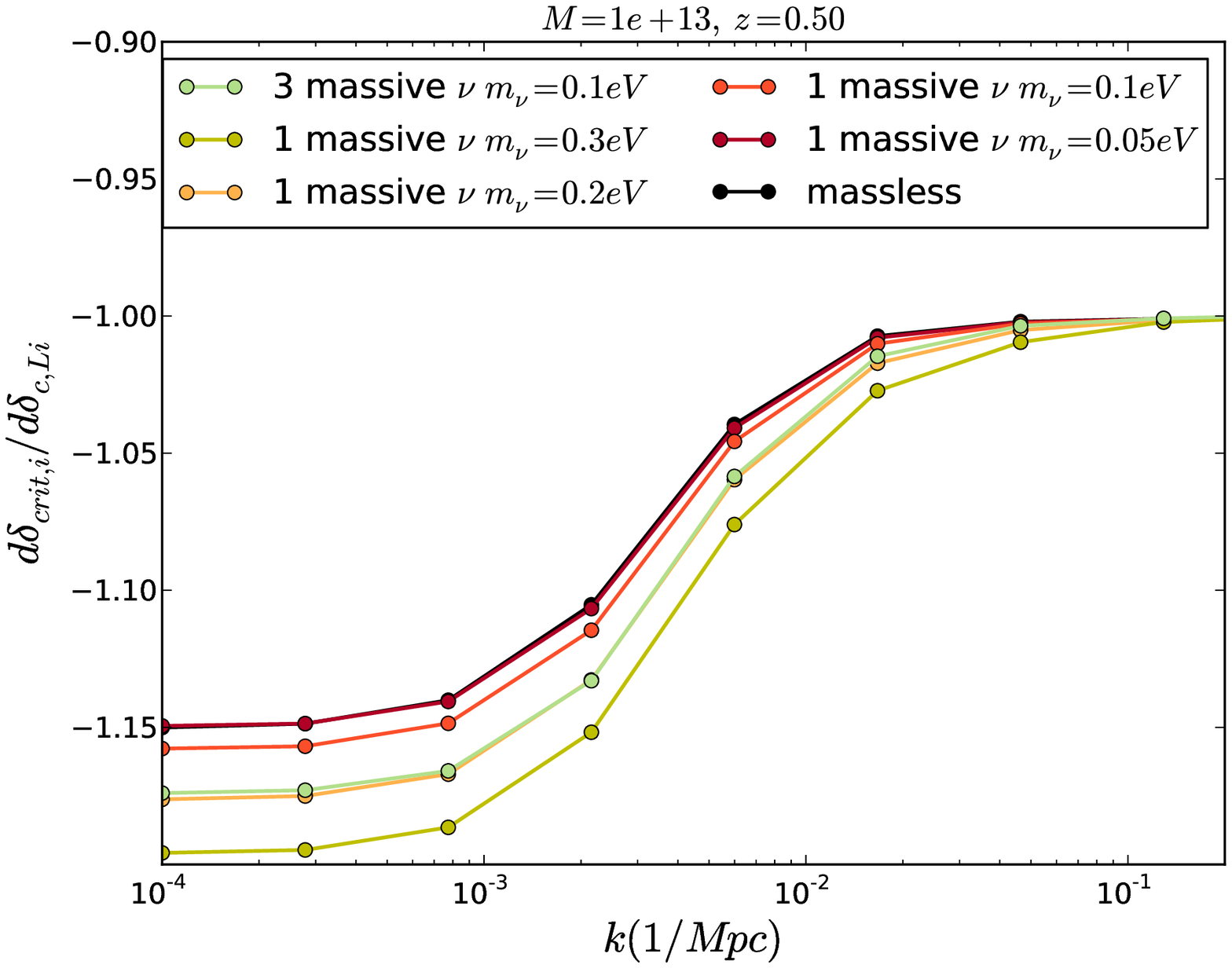} & \includegraphics[width=0.5\textwidth]{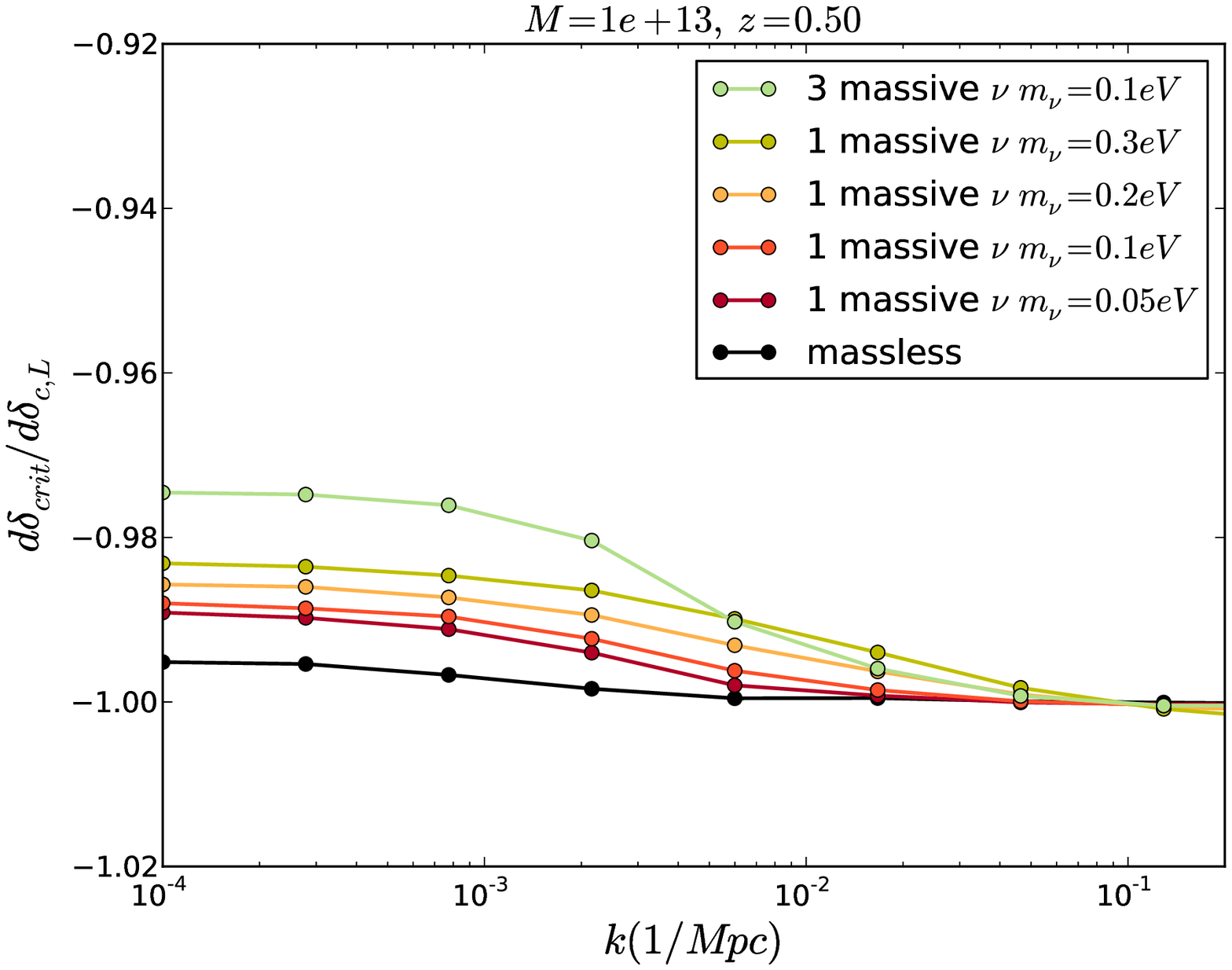}
\end{array}$
\caption{\label{fig:dddcs} The slopes of the relationships plotted in Fig.~(\ref{fig:deltaczSdeltaczL}) -- that is the slope of the line relating $\delta_{c,S}$ to $\delta_{c,L}$. The left panel plots the relationship between the initial values of the two quantities. The right panel plots it with $\delta_{c,S}$ and $\delta_{c,L}$ evaluated at $z_{collapse}$ (using the scale-dependent linear growth functions). Each curve has a fixed $\Omega_m$, but varying $\Omega_c$ and $\Omega_\nu \approx \sum_i m_{\nu i}$ where the neutrino masses are listed in the plot legend.} 
\end{center}
\end{figure}

In Fig. \ref{fig:dddcs} we plot the slopes of the lines in Fig. \ref{fig:deltaciSdeltaciL}  and Fig. \ref{fig:deltaczSdeltaczL}  as a function of $k$. There is clearly a $k$-dependent feature in the slope of the relation between the values of $\delta_{crit, i}$ and $\delta_{c,iL}$ at the initial time.  The scale dependence of $d\delta_{crit, i}/d\delta_{c,iL}$ is present in cosmologies with massive and massless neutrinos but the amplitude of the difference between $d\delta_{crit, i}/d\delta_{c,iL}$ at low and high $k$ increases with increasing neutrino mass. In panel (b) of the same figure we plot the slopes of the lines relating the values of $\delta_{crit}$ and $\delta_{c,L}$ linearly extrapolated to the collapse time. The scale dependence of $d\delta_{crit}/d\delta_{c,L}$ for the linearly extrapolated quantities is smaller, but still present and this scale dependence will lead to scale dependence in the Lagrangian bias factor. 
 

\section{Scale-dependent bias factors}
\label{sec:numericalbias}
\begin{figure}[t]
\begin{center}
$\begin{array}{cc}
 \includegraphics[width=0.5\textwidth]{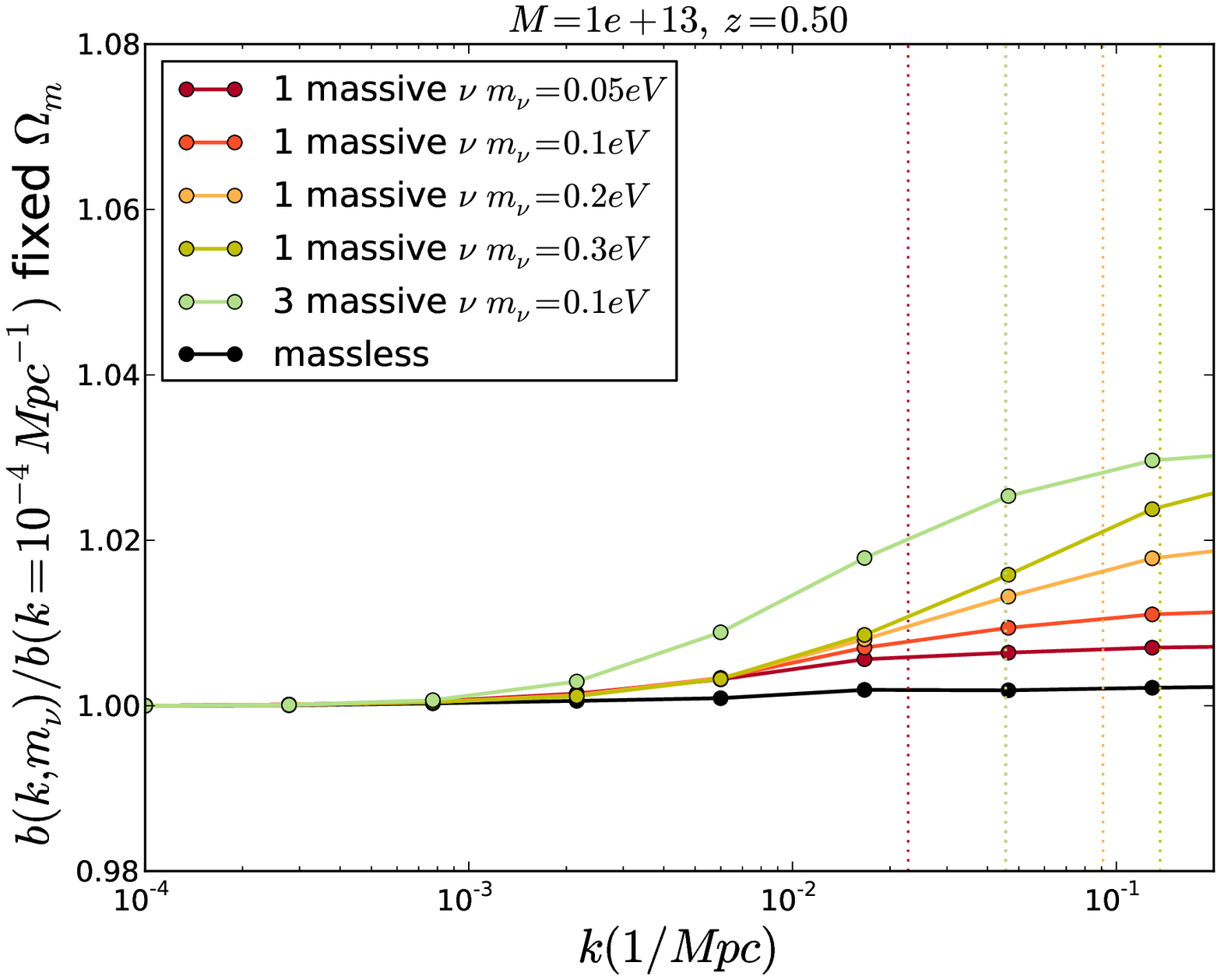} & \includegraphics[width=0.5\textwidth]{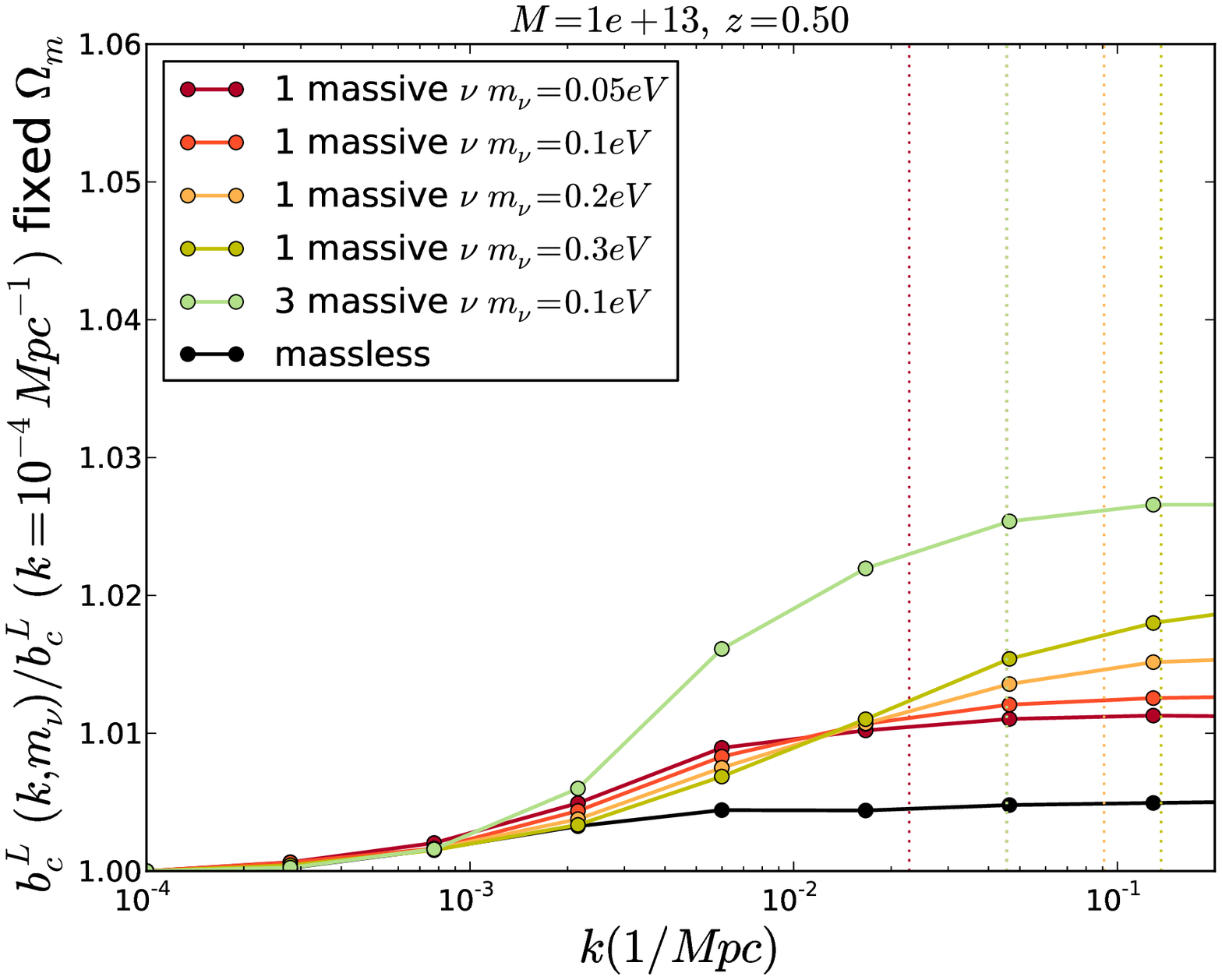}\\
 \includegraphics[width=0.5\textwidth]{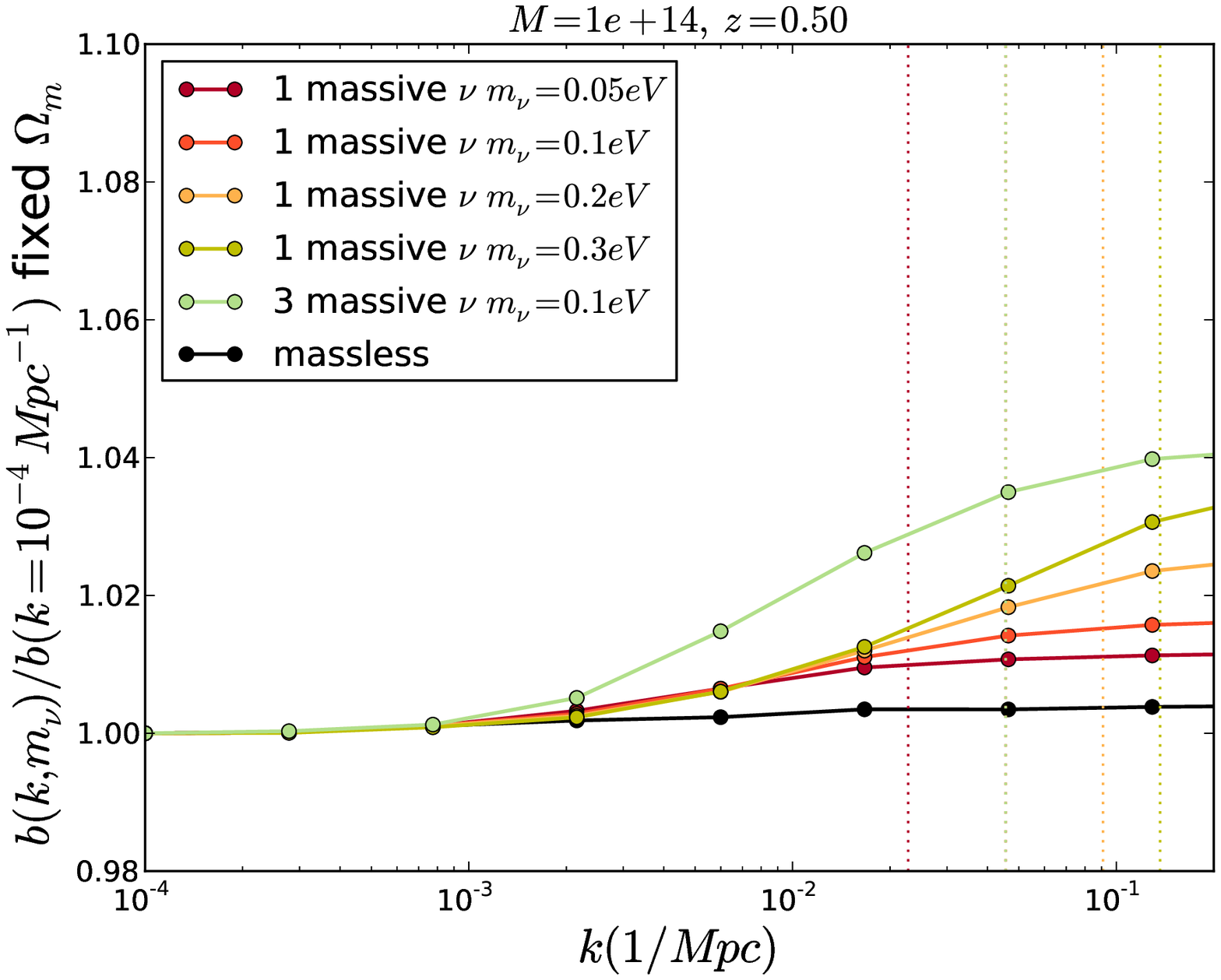} & \includegraphics[width=0.5\textwidth]{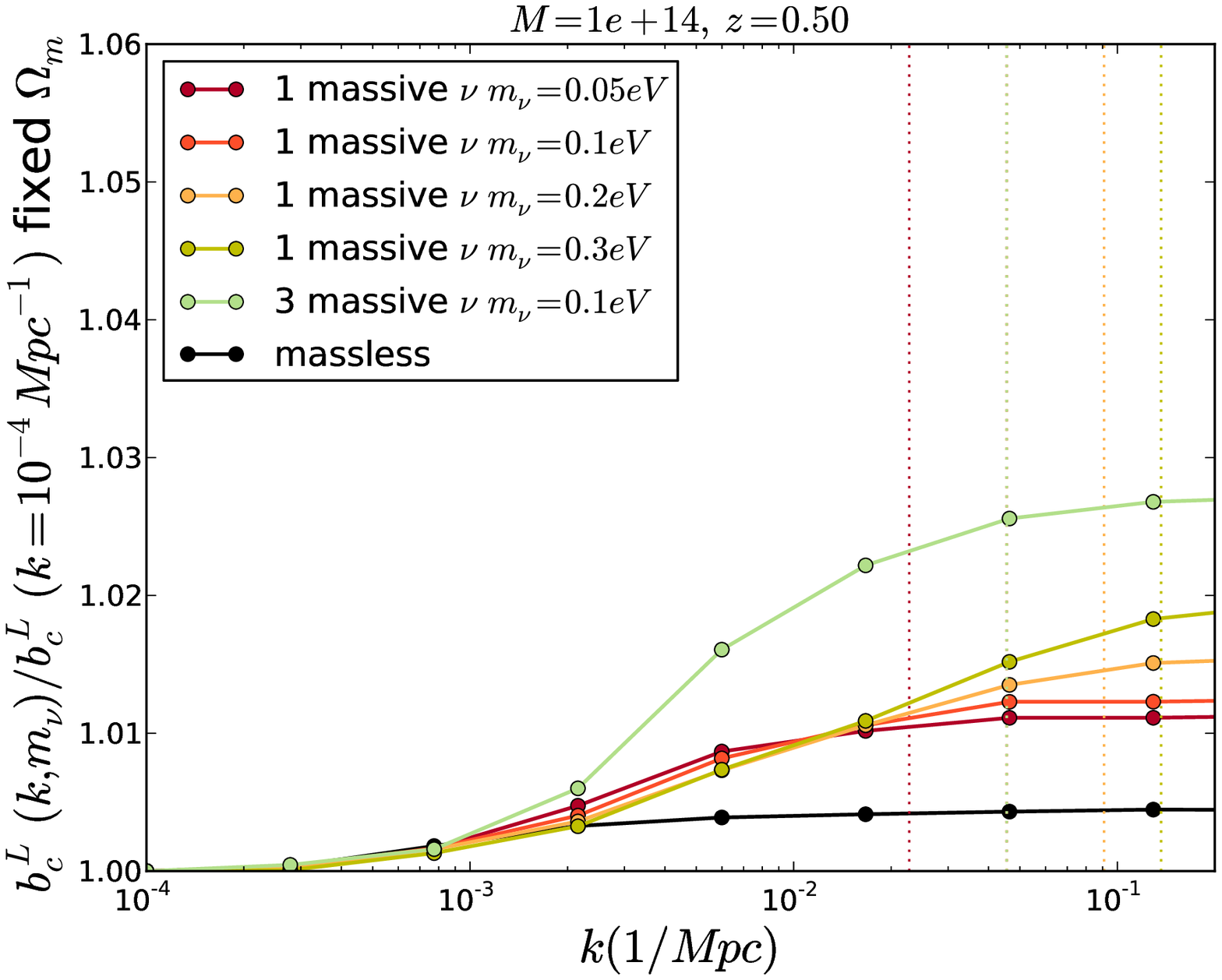}
\end{array}$
\caption{\label{fig:biasesM} The shift in the Eulerian bias (left column) and Lagrangian bias with respect to the CDM (right column) relative to the values of the bias factors at very large scales.  Precisely, the plotted quantity is $b(k)/b(k = 10^{-4}  Mpc^{-1})$. The top row is $b(M)$ for $M = 10^{13} M_\odot$ halos and the bottom row shows the shift in the bias for $M  = 10^{14} M_\odot$ halos.  In all plots the value of $\Omega_m$ is fixed, but $\Omega_c $ and $ \Omega_\nu$ vary. The neutrino free-streaming scale for each hierarchy, Eq.~(\ref{eq:kfs}), is shown by the vertical dotted lines of the same color.  In both panels the order of the legend matches the order of the curves. }
\end{center}
\end{figure}

The calculations of  $d\delta_{crit}/d\delta_{c,L}$ from the previous section along with an expression for the halo mass function allow us to calculate the scale-dependent halo bias in the presence of massive neutrinos. The halo mass function of \cite{Bhattacharya:2010wy} gives
\be
\frac{d\ln n}{d\delta_{crit}} = \frac{q - a(\delta_{crit}/\sigma)^2 }{\delta_{crit}} - \frac{2p/\delta_{crit}}{1+(a\delta_{crit}^2/\sigma^2)^p}
\ee
where $q = 1.795$, $a = 0.788/(1+z)^{0.01}$, and $p = 0.807$. The Lagrangian bias with respect to the CDM is then
\be
\label{eq:bLofk}
b_c^{Lagrangian}(k, m_{\nu}) = \frac{d\ln n}{d\delta_{crit}} \frac{d\delta_{crit}}{d\delta_{c,L}}(k)
\ee
and the Eulerian bias is
\be
\label{eq:bofk}
b(k, m_{\nu}) = \left(1 +\frac{d\ln n}{d\delta_{crit}} \frac{d\delta_{crit}}{d\delta_{c,L}}(k)\right)\frac{f_cP_{cc}(k) + f_\nu P_{c\nu}(k)}{P_{mm}(k)}\,.
\ee
Numerical results for the scale-dependent Eulerian and Lagrangian biases are plotted for halos of mass $M = 10^{13}M_\odot$ and $M = 10^{14}M_\odot$ in Fig. \ref{fig:biasesM}. The neutrino feature is a visible step in the halo bias around the neutrino free-streaming scale (for neutrino mass hierarchies that are not degenerate we've shown the free-streaming scale defined by the most massive neutrino eigenstate). The size of the neutrino step is larger for more massive halos and increases with increasing $\Omega_\nu$. For neutrino mass hierarchies with common $\Omega_\nu$ but different individual $m_{\nu i}$ the bias feature is similar but clearly distinguishable. For instance, comparing the scenarios with $m_{\nu 1} =0.3\,eV$, $m_{\nu 2} = m_{\nu 3} = 0 \,eV$, and $m_{\nu i} = 0.1\,eV$ in Fig. \ref{fig:biasesM} we can see that for three degenerate neutrino mass eigenstates the amplitude of the feature is larger and shifted to larger scales than the feature in the bias for a single massive neutrino with $m_\nu = \sum_i m_{\nu i}$. 

\begin{figure}[t]
\begin{center}
$\begin{array}{cc}
 \includegraphics[width=0.5\textwidth]{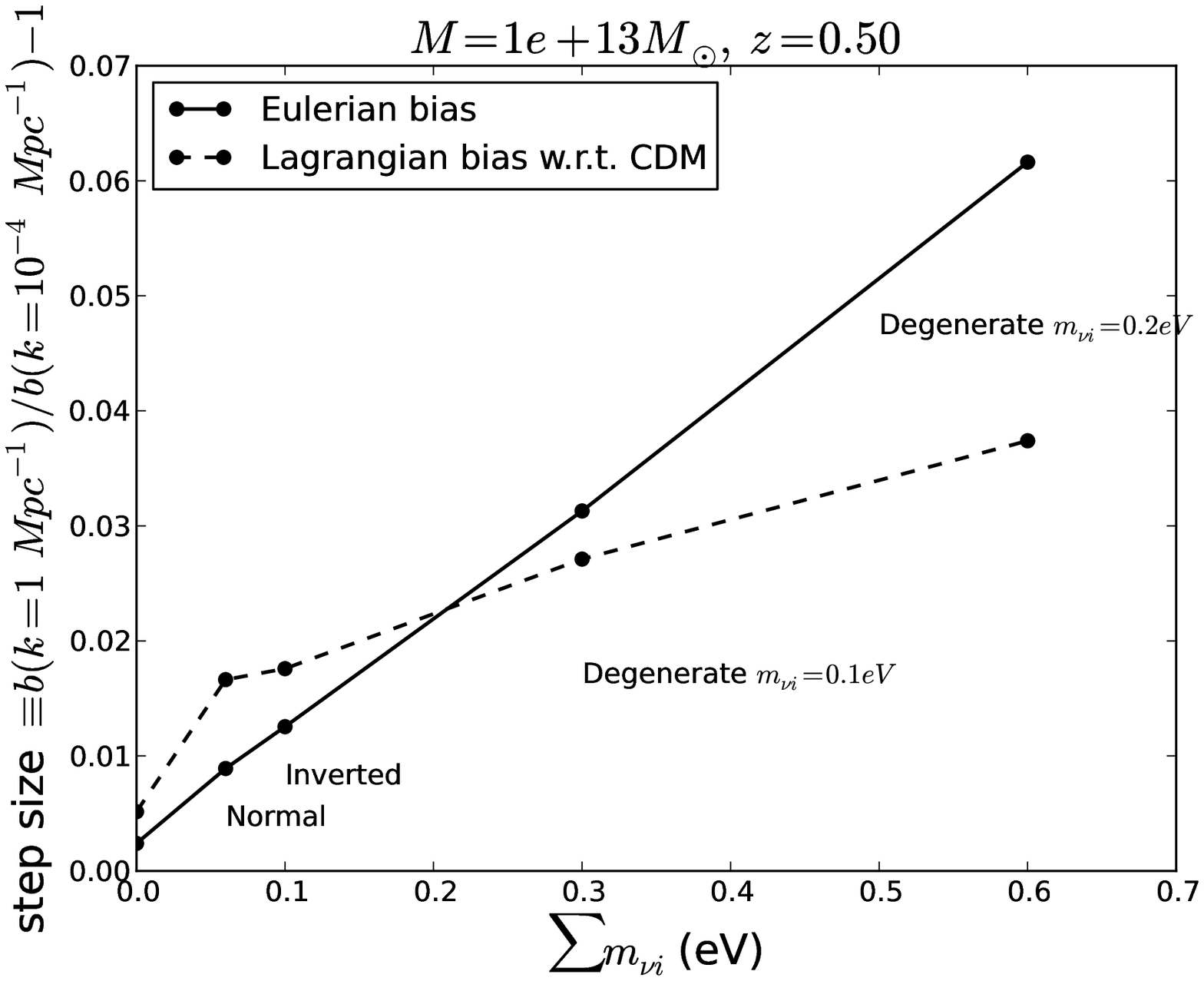} & \includegraphics[width=0.5\textwidth]{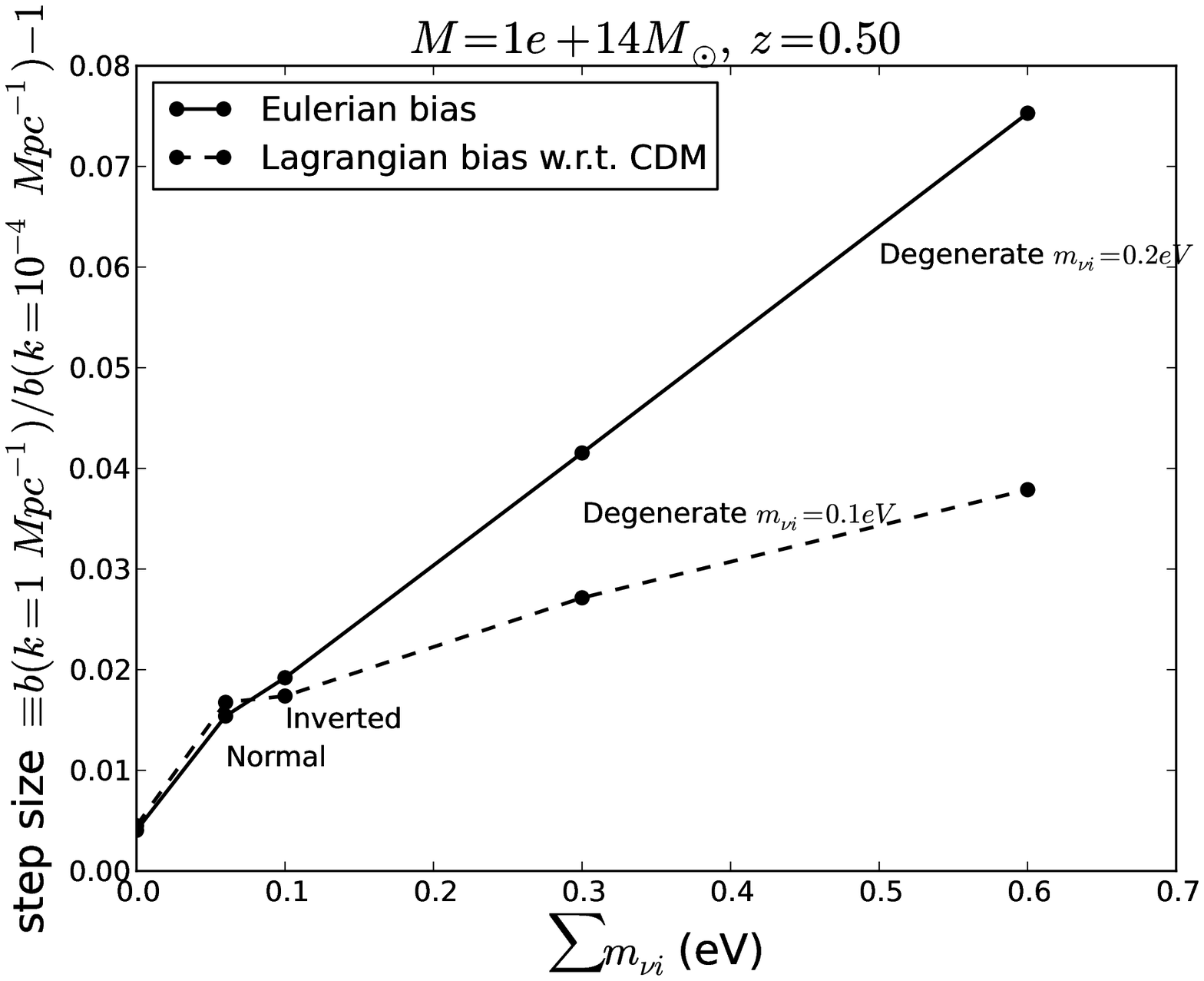}\\
\end{array}$
\caption{\label{fig:stepsize} The amplitude of the step feature in the halo bias seen in Fig.~\ref{fig:biasesM} defined here as $b(k = 1 Mpc^{-1})/b(k = 10^{-4} Mpc^{-1}) -1$. Plotted is the size of the step in the Eulerian bias (solid lines) and Lagrangian bias with respect to the CDM fluctuations (dashed lines) for a range of neutrino mass hierarchies compatible with oscillation data. }
\end{center}
\end{figure}

The sensitivity of the neutrino step in the bias to $\Omega_\nu$ and $M$ is illustrated in Fig. \ref{fig:stepsize}. In that figure we show the size of the step, simply defined as the fractional difference between $b(k)$ at $k = 10^{-4} Mpc^{-1}$ and $k = 1 Mpc^{-1}$, for a range of neutrino mass hierarchies. The step feature is clearly present in both the Lagrangian and Eulerian biases and, as expected from the analytic estimates in \S \ref{sec:halobiasoutline},  the size of the feature increases roughly linearly with increasing total neutrino mass. The fractional step size in the Lagrangian bias with respect to the CDM is nearly independent of $b$ (or $M$). On the other hand, from Eqs.~(\ref{eq:bkestimate}) and (\ref{eq:bLofk}) the step size in the Eulerian halo bias is roughly 
\be
\label{eq:deltabE}
\frac{\Delta b(k)}{b} \approx f_\nu + \frac{b -1}{b}\frac{\Delta b^{Lagrangian}_c}{b^{Lagrangian}_c}
\ee
so the feature in the Eulerian halo bias depends on the population of halos (through $b$) even if $\Delta b_c^{Lagrangian}/b_c^{Lagrangian}$ is independent of halo mass. 

Recall that the neutrino-induced suppression in the linear matter power spectrum is $\sim 8f_\nu$ \cite{Hu:1997mj}. For a scale-independent bias the suppression of the halo or galaxy autopower spectrum is identical to the suppression in the matter power spectrum and the constraints on $f_\nu$ are independent of the population of galaxies.  In cosmologies with massive neutrinos, the halo bias increases on scales below the neutrino free-streaming scale. This increase causes the halo auto-power spectrum to be less suppressed on small scales than one would naively find by assuming a constant bias factor, and from Eq.~(\ref{eq:deltabE}) the amount by which the small-scale power is suppressed depends on $b(M)$. 

\begin{figure}[t]
\begin{center}
$\begin{array}{cc}
 \includegraphics[width=0.5\textwidth]{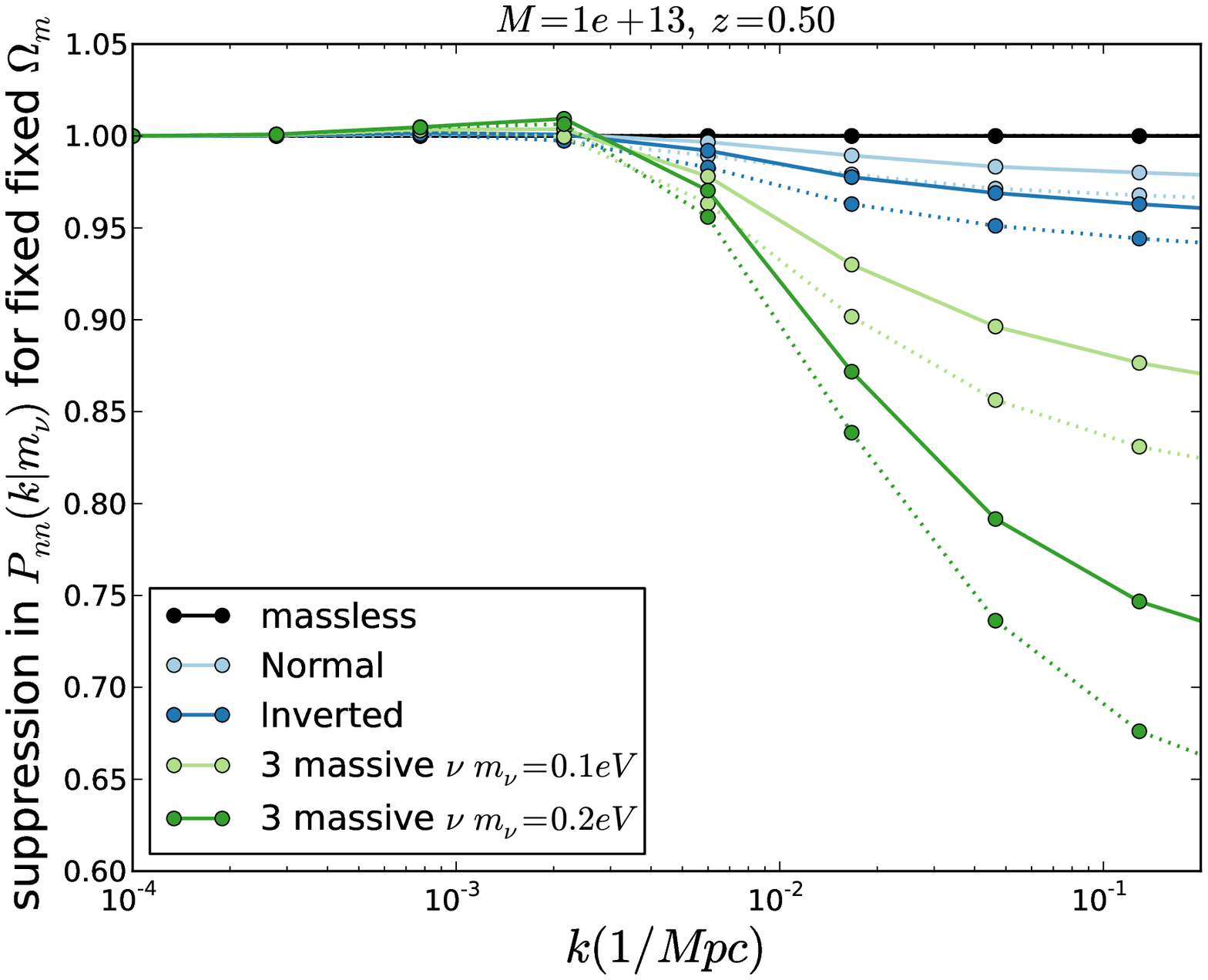} & \includegraphics[width=0.5\textwidth]{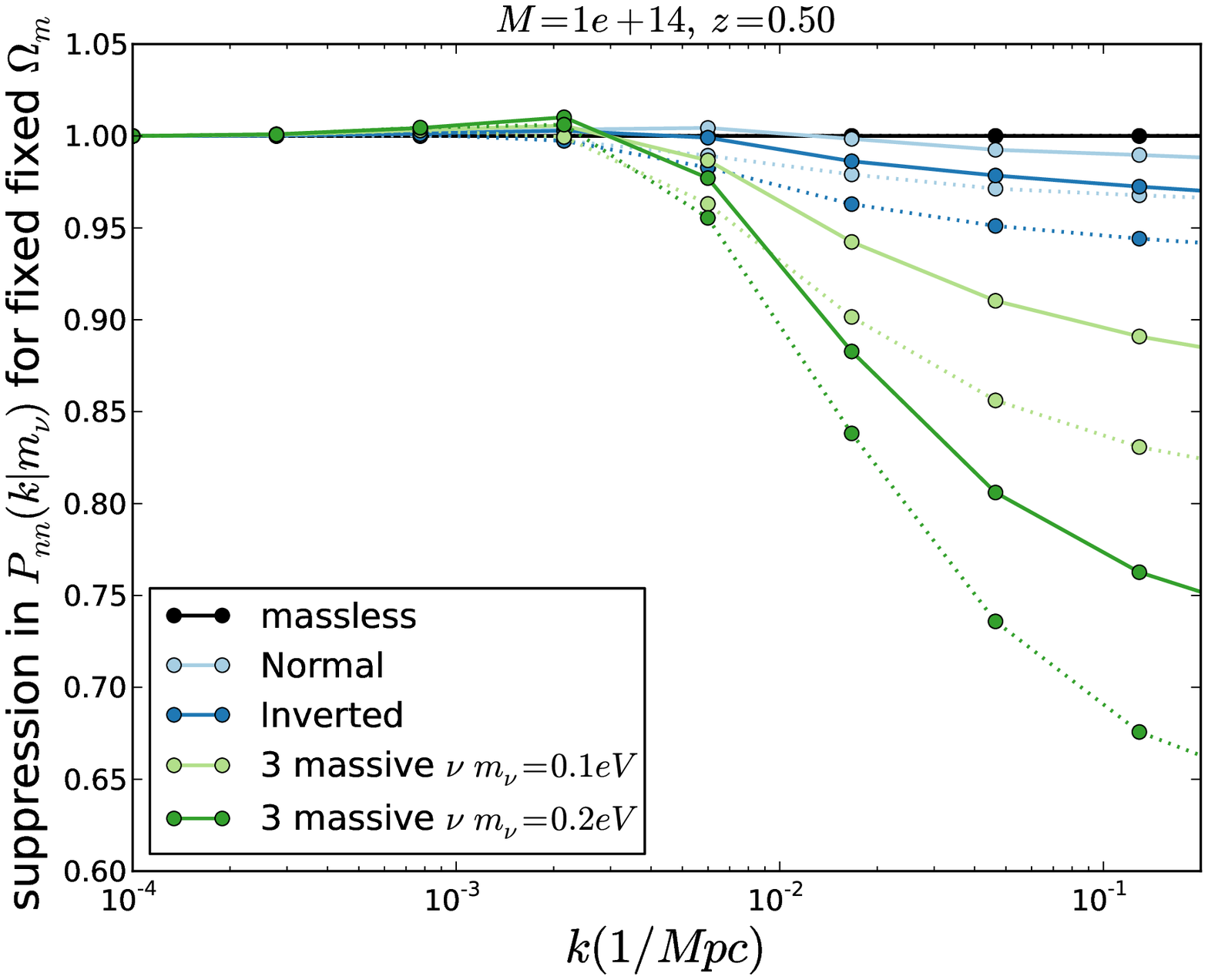}\\
\end{array}$
\caption{\label{fig:Pnn} The suppression in the halo power spectrum in the presence of massive neutrinos, relative to cosmologies without massive neutrinos. Precisely, the plotted quantity is the fractional difference between $P_{nn}(k|m_{\nu})/P_{nn}(k = 10^{-4} Mpc^{-1} | m_{\nu})$ and $P_{nn}(k|m_{\nu} = 0)/P_{nn}(k = 10^{-4} Mpc^{-1} | m_{\nu} =0)$. The solid lines include the scale-dependent bias calculated in this paper, the dotted lines use the standard prediction of a constant value of $b$ defined by  $b = 1 + b_c^{Lagrangian}$ where $b_c^{Lagrangian}$ is calculated assuming $d\delta_{crit}/d\delta_{c,L} = -1$. }
\end{center}
\end{figure}

The change to the halo auto-power spectrum from scale-dependent bias is shown in Fig.~(\ref{fig:Pnn}). The halo auto-power spectrum, including both scale-dependent bias and the scale dependence in $P_{mm}(k)$, is clearly suppressed on small scales and the amount of suppression increases with increasing neutrino mass fraction (solid lines). However the net suppression in $P_{nn}(k)$, including the scale-dependent bias, is smaller than one would have found if a constant bias factor was assumed (dotted lines).  This fact should cause the constraints on neutrino mass from galaxy surveys to relax slightly.  Comparing the two panels in Fig. \ref{fig:stepsize} or \ref{fig:Pnn} one can see that the change to the neutrino mass constraints from scale-dependent halo bias depends on the population of galaxies so we do not attempt to quantify this here. A very rough estimate can, however, be obtained from Eq.~(\ref{eq:deltabE}) and the fact that $\Delta b_c^{Lagrangian}/b_c^{Lagrangian} \sim f_\nu$; in this case the suppression in $P_{nn}$ is reduced from $-8f_\nu$ to $(-6 +2(b-1)/b )f_\nu$ so for a population of galaxies with $b \approx 2$ the sensitivity to $f_\nu$ is decreased by about $40\%$. From our numerical calculations for halos of $M = 10^{13} M_\odot$, $M = 10^{14} M_\odot$ with bias factors roughly $b \sim 1$, $b \sim 2$ respectively, we find that the suppression in the halo auto-power spectra is decreased by $\sim 30\%$ relative to the matter power spectrum. For comparable populations of halos, the constraints on $f_\nu$ from the suppression in the halo autopower spectrum would be expected to relax by a similar amount.

\section{conclusion}
\label{sec:conclusions}
In this paper we have studied halo bias in cosmologies with massive neutrinos and cold dark matter. To do this, we developed a simple framework for calculating the Lagrangian bias from spherical collapse on a long-wavelength mode. The change to the local collapse threshold in the presence of a long-wavelength mode, together with an analytic expression for the mass function, gives the linear halo bias in cosmologies with massive neutrinos. In our calculations we have assumed that the fluctuations in the energy density are adiabatic, that is, we have assumed that the fluctuations in the energy density of different components are coherent. An important extension of these calculations would be to repeat the calculations here in the presence of isocurvature perturbations. 

Interestingly, we find that the halo bias is scale dependent. In cosmologies with massive neutrinos there is a small scale-dependent step in the halo bias around the neutrino free-streaming scale (see Fig.~\ref{fig:biasesM}). The amplitude of the feature is larger for more massive halos and in cosmologies with larger $\Omega_\nu$. We further find that even in a cosmology with massless neutrinos the halo bias is not precisely scale invariant; in this case there is a tiny feature around the matter radiation equality scale. Part of the scale-dependent bias studied here can be understood in terms of the scale-dependent growth of fluctuations in the matter density and is similar to the analysis of \cite{Parfrey:2010uy}. In \cite{Parfrey:2010uy} scale-dependent halo bias arises from scale-dependent growth associated with the late-time accelerated expansion of the universe. In our case, the growth of perturbations is scale dependent at earlier times due to the presence of massive neutrinos, and to a small extent radiation. 

Interestingly, scale-dependent bias has been seen in the recent neutrino-CDM simulations of \cite{Villaescusa-Navarro:2013pva,Castorina:2013wga}. In those works, the authors find a suppression in the halo bias which, from Fig. 6 of \cite{Castorina:2013wga}, appears to be in excellent agreement with our calculations over the same range of scales. Moreover, the authors find that the scale-dependent bias is reduced when they consider the bias with respect to the CDM fluctuations only, $b_c(k) \equiv P_{nc}/P_{cc}$. For the bias defined with respect to the CDM only, we predict $b_c(k) = 1 + b_c^{Lagrangian}(k |m_{\nu})$. This bias factor is still scale dependent but the magnitude of the scale-dependent feature is considerably smaller than in $b(k) = P_{nm}(k)/P_{mm}(k) \approx b_c(k) P_{cm}(k)/P_{mm}(k)$. 
Finally, the scale-dependent halo bias predicted here increases on scales smaller than the neutrino free-streaming length. This means that the suppression in the galaxy power spectrum on scales below $k_{fs}$ is reduced, diminishing the sensitivity of the galaxy power spectrum to $\Omega_\nu$ (see Fig. \ref{fig:Pnn}). We leave the examination of precisely how this effect alters the constraints on neutrino mass from galaxy surveys to further study. 

Finally, the existence of a scale-dependent bias feature offers the opportunity for a new method for constraining neutrino mass through the measurement of the location and/or amplitude of the neutrino feature in the bias. We explore the possibility of constraining neutrino mass from the scale-dependent halo bias in a separate paper \cite{Loverde2014b}. 

\acknowledgements
M.L. thanks Neal Dalal, Daniel Grin, Wayne Hu, and Doug Rudd for many helpful discussions and Matias Zaldarriaga for many helpful discussions and initial collaboration on this project. M.L. is grateful for hospitality at the Institute for Advanced Study while this work was being completed. M.L. is supported by U.S. Department of Energy Award No. DE-FG02-13ER41958.

\bibliographystyle{ieeetr}
\bibliography{mdm}

\begin{thebibliography}{10}

\bibitem{Bond:1980ha}
J.~Bond, G.~Efstathiou, and J.~Silk, ``{Massive Neutrinos and the Large Scale
  Structure of the Universe},'' {\em Phys.Rev.Lett.}, vol.~45, pp.~1980--1984,
  1980.

\bibitem{Hu:1997mj}
W.~Hu, D.~J. Eisenstein, and M.~Tegmark, ``{Weighing neutrinos with galaxy
  surveys},'' {\em Phys.Rev.Lett.}, vol.~80, pp.~5255--5258, 1998.

\bibitem{Seljak:2006bg}
U.~Seljak, A.~Slosar, and P.~McDonald, ``{Cosmological parameters from
  combining the Lyman-alpha forest with CMB, galaxy clustering and SN
  constraints},'' {\em JCAP}, vol.~0610, p.~014, 2006.

\bibitem{Saito:2009ah}
S.~Saito, M.~Takada, and A.~Taruya, ``{Nonlinear power spectrum in the presence
  of massive neutrinos: perturbation theory approach, galaxy bias and parameter
  forecasts},'' {\em Phys.Rev.}, vol.~D80, p.~083528, 2009.

\bibitem{Reid:2009nq}
B.~A. Reid, L.~Verde, R.~Jimenez, and O.~Mena, ``{Robust Neutrino Constraints
  by Combining Low Redshift Observations with the CMB},'' {\em JCAP},
  vol.~1001, p.~003, 2010.

\bibitem{Thomas:2009ae}
S.~A. Thomas, F.~B. Abdalla, and O.~Lahav, ``{Upper Bound of 0.28eV on the
  Neutrino Masses from the Largest Photometric Redshift Survey},'' {\em
  Phys.Rev.Lett.}, vol.~105, p.~031301, 2010.

\bibitem{Swanson:2010sk}
M.~E. Swanson, W.~J. Percival, and O.~Lahav, ``{Neutrino Masses from Clustering
  of Red and Blue Galaxies: A Test of Astrophysical Uncertainties},'' {\em
  Mon.Not.Roy.Astron.Soc.}, vol.~409, pp.~1100--1112, 2010.

\bibitem{Xia:2012na}
J.-Q. Xia, B.~R. Granett, M.~Viel, S.~Bird, L.~Guzzo, {\em et~al.},
  ``{Constraints on Massive Neutrinos from the CFHTLS Angular Power
  Spectrum},'' {\em JCAP}, vol.~1206, p.~010, 2012.

\bibitem{RiemerSorensen:2011fe}
S.~Riemer-Sorensen, C.~Blake, D.~Parkinson, T.~M. Davis, S.~Brough, {\em
  et~al.}, ``{The WiggleZ Dark Energy Survey: Cosmological neutrino mass
  constraint from blue high-redshift galaxies},'' {\em Phys.Rev.}, vol.~D85,
  p.~081101, 2012.

\bibitem{Zhao:2012xw}
G.-B. Zhao, S.~Saito, W.~J. Percival, A.~J. Ross, F.~Montesano, {\em et~al.},
  ``{The clustering of galaxies in the SDSS-III Baryon Oscillation
  Spectroscopic Survey: weighing the neutrino mass using the galaxy power
  spectrum of the CMASS sample},'' 2012.

\bibitem{dePutter:2012sh}
R.~de~Putter, O.~Mena, E.~Giusarma, S.~Ho, A.~Cuesta, {\em et~al.}, ``{New
  Neutrino Mass Bounds from Sloan Digital Sky Survey III Data Release 8
  Photometric Luminous Galaxies},'' {\em Astrophys.J.}, vol.~761, p.~12, 2012.

\bibitem{Lesgourgues:2006nd}
J.~Lesgourgues and S.~Pastor, ``{Massive neutrinos and cosmology},'' {\em
  Phys.Rept.}, vol.~429, pp.~307--379, 2006.

\bibitem{Ade:2013zuv}
P.~Ade {\em et~al.}, ``{Planck 2013 results. XVI. Cosmological parameters},''
  {\em Astron.Astrophys.}, 2014.

\bibitem{Hinshaw:2012aka}
G.~Hinshaw {\em et~al.}, ``{Nine-Year Wilkinson Microwave Anisotropy Probe
  (WMAP) Observations: Cosmological Parameter Results},'' 2012.

\bibitem{Hou:2012xq}
Z.~Hou, C.~Reichardt, K.~Story, B.~Follin, R.~Keisler, {\em et~al.},
  ``{Constraints on Cosmology from the Cosmic Microwave Background Power
  Spectrum of the 2500-square degree SPT-SZ Survey},'' 2012.

\bibitem{Sievers:2013ica}
J.~L. Sievers, R.~A. Hlozek, M.~R. Nolta, V.~Acquaviva, G.~E. Addison, {\em
  et~al.}, ``{The Atacama Cosmology Telescope: Cosmological parameters from
  three seasons of data},'' 2013.

\bibitem{Beringer:1900zz}
J.~Beringer {\em et~al.}, ``{Review of Particle Physics (RPP)},'' {\em
  Phys.Rev.}, vol.~D86, p.~010001, 2012.

\bibitem{Saito2010}
S.~{Saito}, M.~{Takada}, and A.~{Taruya}, ``{Neutrino mass constraint from the
  Sloan Digital Sky Survey power spectrum of luminous red galaxies and
  perturbation theory},'' {\em Phys.Rev.}, vol.~D83, Feb. 2011.

\bibitem{Beutler:2013yhm}
F.~Beutler {\em et~al.}, ``{The clustering of galaxies in the SDSS-III Baryon
  Oscillation Spectroscopic Survey: Testing gravity with redshift-space
  distortions using the power spectrum multipoles},'' 2013.

\bibitem{Vikhlinin:2008ym}
A.~Vikhlinin, A.~Kravtsov, R.~Burenin, H.~Ebeling, W.~Forman, {\em et~al.},
  ``{Chandra Cluster Cosmology Project III: Cosmological Parameter
  Constraints},'' {\em Astrophys.J.}, vol.~692, pp.~1060--1074, 2009.

\bibitem{Mantz:2009rj}
A.~Mantz, S.~W. Allen, and D.~Rapetti, ``{The Observed Growth of Massive Galaxy
  Clusters IV: Robust Constraints on Neutrino Properties},'' {\em
  Mon.Not.Roy.Astron.Soc.}, vol.~406, pp.~1805--1814, 2010.

\bibitem{Benson:2011uta}
B.~Benson, T.~de~Haan, J.~Dudley, C.~Reichardt, K.~Aird, {\em et~al.},
  ``{Cosmological Constraints from Sunyaev-Zel'dovich-Selected Clusters with
  X-ray Observations in the First 178 Square Degrees of the South Pole
  Telescope Survey},'' {\em Astrophys.J.}, vol.~763, p.~147, 2013.

\bibitem{Reichardt:2012yj}
C.~Reichardt, B.~Stalder, L.~Bleem, T.~Montroy, K.~Aird, {\em et~al.},
  ``{Galaxy clusters discovered via the Sunyaev-Zel'dovich effect in the first
  720 square degrees of the South Pole Telescope survey},'' {\em Astrophys.J.},
  vol.~763, p.~127, 2013.

\bibitem{Hasselfield:2013wf}
M.~Hasselfield, M.~Hilton, T.~A. Marriage, G.~E. Addison, L.~F. Barrientos,
  {\em et~al.}, ``{The Atacama Cosmology Telescope: Sunyaev-Zel'dovich selected
  galaxyclusters at 148 GHz from three seasons of data},'' {\em JCAP},
  vol.~1307, p.~008, 2013.

\bibitem{Ade:2013lmv}
P.~Ade {\em et~al.}, ``{Planck 2013 results. XX. Cosmology from
  Sunyaev-Zeldovich cluster counts},'' 2013.

\bibitem{Hui:2007zh}
L.~Hui and K.~P. Parfrey, ``{The Evolution of Bias: Generalized},'' {\em
  Phys.Rev.}, vol.~D77, p.~043527, 2008.

\bibitem{Parfrey:2010uy}
K.~Parfrey, L.~Hui, and R.~K. Sheth, ``{Scale-dependent halo bias from
  scale-dependent growth},'' {\em Phys.Rev.}, vol.~D83, p.~063511, 2011.

\bibitem{Gunn:1972sv}
J.~E. Gunn and I.~Gott, J.~Richard, ``{On the Infall of Matter into Clusters of
  Galaxies and Some Effects on Their Evolution},'' {\em Astrophys.J.},
  vol.~176, pp.~1--19, 1972.

\bibitem{Sheth:1999mn}
R.~K. Sheth and G.~Tormen, ``{Large scale bias and the peak background
  split},'' {\em Mon.Not.Roy.Astron.Soc.}, vol.~308, p.~119, 1999.

\bibitem{Cole:1989vx}
S.~Cole and N.~Kaiser, ``{Biased clustering in the cold dark matter
  cosmogony},'' {\em Mon.Not.Roy.Astron.Soc.}, vol.~237, pp.~1127--1146, 1989.

\bibitem{Manera:2009ak}
M.~Manera, R.~K. Sheth, and R.~Scoccimarro, ``{Large scale bias and the
  inaccuracy of the peak-background split},'' 2009.

\bibitem{LoVerde:2013lta}
M.~LoVerde and M.~Zaldarriaga, ``{Neutrino clustering around spherical dark
  matter halos},'' {\em Phys.Rev.}, vol.~D89, p.~063502, 2014.

\bibitem{LoVerde:2014rxa}
M.~LoVerde, ``{Spherical collapse in $\nu \Lambda CDM$},'' 2014.

\bibitem{Colin:2007bk}
P.~Colin, O.~Valenzuela, and V.~Avila-Reese, ``{On the Structure of Dark Matter
  Halos at the Damping Scale of the Power Spectrum with and without Relict
  Velocities},'' {\em Astrophys.J.}, vol.~673, pp.~203--214, 2008.

\bibitem{Brandbyge:2008rv}
J.~Brandbyge, S.~Hannestad, T.~Haugbolle, and B.~Thomsen, ``{The Effect of
  Thermal Neutrino Motion on the Non-linear Cosmological Matter Power
  Spectrum},'' {\em JCAP}, vol.~0808, p.~020, 2008.

\bibitem{Agarwal:2010mt}
S.~Agarwal and H.~A. Feldman, ``{The effect of massive neutrinos on the matter
  power spectrum},'' {\em Mon.Not.Roy.Astron.Soc.}, vol.~410, p.~1647, 2011.

\bibitem{Viel:2010bn}
M.~Viel, M.~G. Haehnelt, and V.~Springel, ``{The effect of neutrinos on the
  matter distribution as probed by the Intergalactic Medium},'' {\em JCAP},
  vol.~1006, p.~015, 2010.

\bibitem{Brandbyge:2010ge}
J.~Brandbyge, S.~Hannestad, T.~Haugboelle, and Y.~Y. Wong, ``{Neutrinos in
  Non-linear Structure Formation - The Effect on Halo Properties},'' {\em
  JCAP}, vol.~1009, p.~014, 2010.

\bibitem{Marulli:2011he}
F.~Marulli, C.~Carbone, M.~Viel, L.~Moscardini, and A.~Cimatti, ``{Effects of
  Massive Neutrinos on the Large-Scale Structure of the Universe},'' {\em
  Mon.Not.Roy.Astron.Soc.}, vol.~418, p.~346, 2011.

\bibitem{VillaescusaNavarro:2012ag}
F.~Villaescusa-Navarro, S.~Bird, C.~Pena-Garay, and M.~Viel, ``{Non-linear
  evolution of the cosmic neutrino background},'' {\em JCAP}, vol.~1303,
  p.~019, 2013.

\bibitem{Upadhye:2013ndm}
A.~Upadhye, R.~Biswas, A.~Pope, K.~Heitmann, S.~Habib, {\em et~al.},
  ``{Large-Scale Structure Formation with Massive Neutrinos and Dynamical Dark
  Energy},'' 2013.

\bibitem{Villaescusa-Navarro:2013pva}
F.~Villaescusa-Navarro, F.~Marulli, M.~Viel, E.~Branchini, E.~Castorina, {\em
  et~al.}, ``{Cosmology with massive neutrinos I: towards a realistic modeling
  of the relation between matter, haloes and galaxies},'' 2013.

\bibitem{Castorina:2013wga}
E.~Castorina, E.~Sefusatti, R.~K. Sheth, F.~Villaescusa-Navarro, and M.~Viel,
  ``{Cosmology with massive neutrinos II: on the universality of the halo mass
  function and bias},'' {\em JCAP}, vol.~1402, p.~049, 2014.

\bibitem{Costanzi:2013bha}
M.~Costanzi, F.~Villaescusa-Navarro, M.~Viel, J.-Q. Xia, S.~Borgani, {\em
  et~al.}, ``{Cosmology with massive neutrinos III: the halo mass function
  andan application to galaxy clusters},'' {\em JCAP}, vol.~1312, p.~012, 2013.

\bibitem{Smith:2006ne}
R.~E. Smith, R.~Scoccimarro, and R.~K. Sheth, ``{The Scale Dependence of Halo
  and Galaxy Bias: Effects in Real Space},'' {\em Phys.Rev.}, vol.~D75,
  p.~063512, 2007.

\bibitem{Matsubara:2008wx}
T.~Matsubara, ``{Nonlinear perturbation theory with halo bias and
  redshift-space distortions via the Lagrangian picture},'' {\em Phys.Rev.},
  vol.~D78, p.~083519, 2008.

\bibitem{McDonald:2008}
P.~{McDonald} and A.~{Roy}, ``{Clustering of dark matter tracers: generalizing
  bias for the coming era of precision LSS},'' vol.~8, p.~20, Aug. 2009.

\bibitem{Desjacques2010}
V.~{Desjacques}, M.~{Crocce}, R.~{Scoccimarro}, and R.~K. {Sheth}, ``{Modeling
  scale-dependent bias on the baryonic acoustic scale with the statistics of
  peaks of Gaussian random fields},'' {\em Phys.Rev.}, p.~103529, Nov. 2010.

\bibitem{Carlson:2012}
J.~{Carlson}, B.~{Reid}, and M.~{White}, ``{Convolution Lagrangian perturbation
  theory for biased tracers},'' {\em Mon.Not.Roy.Astron.Soc.}, vol.~429,
  pp.~1674--1685, Feb. 2013.

\bibitem{Sheth:2012}
R.~K. {Sheth}, K.~C. {Chan}, and R.~{Scoccimarro}, ``{Nonlocal Lagrangian
  bias},'' {\em Phys.Rev.}, p.~083002, Apr. 2013.

\bibitem{Chan2012}
K.~C. {Chan}, R.~{Scoccimarro}, and R.~K. {Sheth}, ``{Gravity and large-scale
  nonlocal bias},'' {\em Phys.Rev.}, p.~083509, Apr. 2012.

\bibitem{Baldauf:2013hka}
T.~Baldauf, U.~Seljak, R.~E. Smith, N.~Hamaus, and V.~Desjacques, ``{Halo
  Stochasticity from Exclusion and non-linear Clustering},'' {\em Phys.Rev.},
  vol.~D88, p.~083507, 2013.

\bibitem{Tassev:2013zua}
S.~Tassev, ``{N-point Statistics of Large-Scale Structure in the Zel'dovich
  Approximation},'' 2013.

\bibitem{Saito:2014qha}
S.~Saito, T.~Baldauf, Z.~Vlah, U.~Seljak, T.~Okumura, {\em et~al.},
  ``{Understanding higher-order nonlocal halo bias at large scales by combining
  the power spectrum with the bispectrum},'' 2014.

\bibitem{Biagetti:2014pha}
M.~Biagetti, V.~Desjacques, A.~Kehagias, and A.~Riotto, ``{Non-local halo bias
  with and without massive neutrinos},'' 2014.

\bibitem{White1987}
S.~D.~M. {White}, M.~{Davis}, G.~{Efstathiou}, and C.~S. {Frenk}, ``{Galaxy
  distribution in a cold dark matter universe},'' {\em Nature}, vol.~330,
  pp.~451--453, Dec. 1987.

\bibitem{Bardeen:1985tr}
J.~M. Bardeen, J.~Bond, N.~Kaiser, and A.~Szalay, ``{The Statistics of Peaks of
  Gaussian Random Fields},'' {\em Astrophys.J.}, vol.~304, pp.~15--61, 1986.

\bibitem{Lewis:1999bs}
A.~Lewis, A.~Challinor, and A.~Lasenby, ``Efficient computation of {CMB}
  anisotropies in closed {FRW} models,'' {\em Astrophys. J.}, vol.~538,
  pp.~473--476, 2000.

\bibitem{Bhattacharya:2010wy}
S.~Bhattacharya, K.~Heitmann, M.~White, Z.~Lukic, C.~Wagner, {\em et~al.},
  ``{Mass Function Predictions Beyond LCDM},'' {\em Astrophys.J.}, vol.~732,
  p.~122, 2011.

\bibitem{Loverde2014b}
M.~LoVerde, ``{Neutrino mass without cosmic variance?},'' {\em in prep.}

\end{thebibliography}

\end{document}